\documentclass[preprint]{aastex61}

\submitjournal{ApJ}

\shorttitle{Two key parameters controlling 
particle clumping
}
\shortauthors{Sekiya \& Onishi}

\begin{document}

\title{Two key parameters controlling 
particle clumping caused by streaming instability 
in the dead-zone dust layer of a protoplanetary disk}

\correspondingauthor{Minoru Sekiya}

\email{sekiya.minoru.393@m.kyushu-u.ac.jp}

\author{Minoru Sekiya}
\affiliation{Department of Earth and Planetary Sciences, 
Faculty of Science, Kyushu University, \\
744 Motooka, Nishi-ku, Fukuoka 819-0395, Japan}

\author{Isamu K. Onishi}
\affiliation{6-15-2, Kasumigaoka, Higashi-ku, Fukuoka 813-0003, Japan}



\begin{abstract}

The streaming instability and Kelvin--Helmholtz instability are considered the two major sources causing 
clumping of dust particles
and turbulence in the dust layer of a protoplanetary disk as long as we consider the dead zone where the magneto-rotational instability does not grow.
Extensive numerical simulations have been carried out in order to elucidate the condition for the development of 
particle clumping caused by the streaming instability.
In this paper, a set of two parameters suitable  
for classifying the numerical results is proposed.
One is the Stokes number that has been employed in previous works
and the other is the dust particle column density 
that is nondimensionalized using the gas density in the midplane, 
Keplerian angular velocity, 
and difference between the Keplerian and gaseous orbital velocities.
The magnitude of dust clumping is a measure of the behavior of the dust layer.
Using three-dimensional numerical simulations of dust particles and gas
based on Athena code v. 4.2,
it is confirmed that
the magnitude of dust clumping
for two disk models are
 similar
if 
the corresponding sets of values of the two parameters are identical to each other,
even if the values of the metallicity (i.e., the ratio of the columns density of the dust particles to that of the gas) are different.

\end{abstract}




\section{Introduction} \label{sec:1}

Planetesimals are hypothetical small solid bodies that can grow because of their mutual gravity when they collide with each other in a protoplanetary disk. 
It is difficult for silicate particles to stick together 
because of their weak material forces \citep{Blum10};
thus, the most probable way to form planetesimals from millimeter-sized chondrules is the gravitational instability \citep{Safronov69, Goldreich_Ward73, Sekiya83}.
For the gravitational instability to develop, the dust volume density $\rho_d$ (the total mass of dust particles in a unit volume of a disk) must exceed the critical density $\rho_{GI} \sim M_*/r^3$, where
$M_*$ is the mass of the central star and $r$ is the orbital radius.

One of the mechanisms for the dust concentration is the streaming instability
\citep{Youdin_Goodman05, Youdin_Johansen07}.
The dust density grows significantly and may exceed $\rho_{GI}$ in the nonlinear stage of the streaming instability \citep{Johansen_Youdin07, Johansen_etal07, Johansen_etal09,  Simon_etal16, Simon_etal17}.
\citet{Carrera_etal15} thoroughly examined the condition for the formation and growth of 
clumps of dust particles caused by the streaming instability.
\citet{Yang_etal17} revised the condition by performing long-term simulations with higher resolution.

In this paper, a new parameter set is introduced that is 
appropriate for the classification of turbulence and
particle clumping 
in the dead-zone dust layer of a protoplanetary disk.
The magnitude of dust clumping is a measure of the behavior of the dust layer.
Through three-dimensional numerical simulations,
it is shown that 
the magnitude of dust clumping for two disk models are similar if the corresponding sets of values of the two parameters are
identical,
even if the values of the metallicity (i.e., the ratio of the columns density of the dust particles to that of the gas) are different. 
This work will provide a novel criterion for the growth of 
clumps of dust particles caused by the streaming instability. We ignore the effect of self-gravity in this work. The planetesimal formation because of gravitational instability is expected to occur after the volume density of particle clumps formed by the streaming instability exceeds $\rho_{GI}$.

\section{Two key dimensionless parameters} \label{sec:2}

There are two important time scales for the dynamics of the dust layer in a protoplanetary disk.
One is the stopping time of a dust particle $t_{stop}$
and the other is the inverse of the Keplerian angular velocity $\Omega_K^{-1}$.
The ratio of these time scales is
called the Stokes number:
$\tau_s\equiv t_{stop} \Omega_K$, 
which has been used in many previous works as a measure of the frictional coupling of dust particles and gas 
\citep{Youdin_Goodman05, Youdin_Johansen07, Johansen_Youdin07, Carrera_etal15,
 Yang_etal17}.
We also adopt $\tau_s$ as one of the key dimensionless parameters.
In this work, we consider the restriction where $\tau_s < 1$.
This condition is satisfied for any orbital radius if the radius of a dust particle is less than 1 cm and the minimum-mass solar nebula model \citep{Hayashi81} is employed.

There are two characteristic velocities related to the dynamics of the dust layer in a protoplanetary disk.
One is $\eta v_K$, where $v_K=r\Omega_K$ is the Keplerian velocity, $\eta=-(1/2)[(1/\rho_g)(\partial P/\partial r)]/[r\Omega_K^2]$,  
$\rho_g$ is the gas density, 
and $P$ is the gas pressure. 
In usual models of protoplanetary disks, 
the radial pressure gradient is negative. 
In this case, 
$\eta v_K$ represents the delay of the revolution velocity of the gas compared to the Keplerian velocity  \citep{Adachi_etal76, Weidenschilling77}.
Another is the isothermal sound velocity $c_s$ that has two roles for determining the characteristics of a local protoplanetary disk:
(1) The vertical scale height of the gas disk is determined by $H_g=c_s /\Omega_K$;
(2) the dynamical compressibility depends on the sound velocity.
The ratio of these two velocities gives another dimensionless parameter
$\Pi \equiv \eta v_K/c_s$, 
which has also been employed to characterize the numerical results \citep{Bai_Stone10c, Carrera_etal15,
Yang_etal17}. 
Note that $\Pi \ll 1$ for usual disk models.
The linear analysis and the nonlinear simulation of the streaming instability in the unstratified disk \citep{Youdin_Johansen07, Johansen_Youdin07} also employ $\Pi$ as a parameter.
In contrast, the original work on the streaming instability assumes that the gas is incompressible \citep{Youdin_Goodman05}; in this case, $\Pi$ is inadequate for expressing the value of $\eta v_K$, because $c_s = \infty$ and hence $\Pi =0$.
The values of the linear growth rate are similar for both the compressible and incompressible models.
Thus, $\Pi$ may not be an appropriate parameter for characterizing the streaming instability.

It has been revealed by means of numerical simulations 
that the behavior of dust particles and gas depends on the metallicity 
$Z\equiv \Sigma_d/\Sigma_g$, 
where $\Sigma_d$ and $\Sigma_g$ are the column densities of dust particles and gas, respectively 
\citep{Johansen_etal09, Bai_Stone10b, Carrera_etal15, 
Yang_etal17}.
Note that \citet{Carrera_etal15} employed $Z'=\Sigma_d/(\Sigma_g+\Sigma_d)$ instead of Z; however, the difference between $Z$ and $Z'$ is small because $\Sigma_d \ll \Sigma_g$ in the cases under consideration.
If $Z$ exceeds a critical value, particle clumps grow possibly because of the streaming instability \citep{Johansen_etal09}.
The critical value of the metallicity depends on $\Pi$ and $\tau_s$ \citep{Carrera_etal15}.

There are several reasons why the metallicity is usually employed as a parameter.
Historically, only the dust mass could be determined from the observation of a protoplanetary disk, and the gas mass was calculated by assuming the metallicity \citep{Beckwith_etal90}.
Recent observations enable us to determine both the dust and gas masses in protoplanetary disks \citep{Ansdell_etal16}.
The metallicity in the solar protoplanetary disk is determined from the solar abundance of elements in each region where water condenses and vaporizes \citep{Hayashi81}.
Several evolution models of the metallicity 
due to the growth and migration of dust particles as well as the disk photoevaporation have been published 
\citep{Youdin_Shu02, Youdin_Chiang04, Gorti_etal15}.


Although metallicity is important for the models of the protoplanetary disks as mentioned above,
the metallicity is not considered an appropriate parameter 
for characterizing the dynamics of the dust layer.
This is because the dust particles settle toward the midplane and form a thin dust layer.
The scale height of the dust layer is much less than that of the gaseous disk.
It is not likely that gas far from the dust layer affects the dynamics in the dust layer.
Here, we seek a characteristic value that represents the column density of the gas around the dust layer.

First, note that the gas has an almost constant density in the dust layer
if we only consider a local region where the radial width $\Delta r$ is much smaller than the orbital radius $r$.
This is because
the dust layer is much thinner than the vertical scale height of the gas.
Hence, we use the unperturbed midplane gas density $\rho_{g0}$ at $r$ as a representative value.

If the frictional coupling is strong, i.e., $\tau_s \ll 1$, the velocity difference of a dust-rich region where 
$\rho_d \gg \rho_g$
and a gas-rich region where
$\rho_g \gg \rho_d$
 is of the order of $\eta v_K$ \citep{Nakagawa_etal86}. 
The maximum growth rate of the streaming instability and Kelvin--Helmholtz instability is of the order of (but somewhat less than) $\Omega_K$ \citep{Youdin_Goodman05, Chiang08}.
Thus, the length scale appropriate for characterizing the dust layer is 
$\eta v_K/\Omega_K=\eta r$ from the viewpoint of dimensional analysis.
Actually, the thickness of the dust layer when the dust settling and turbulent mixing are in equilibrium is of the order of  $\eta r$ according to the results of numerical simulations by \citet{Carrera_etal15}.

Multiplying $\rho_{g0}$ by $\eta r$, we obtain a gas column density that is appropriate for characterizing the column density of the gas in the dust layer. 
Thus, we find a new dimensionless parameter
\begin{equation}
\sigma_d \equiv \Sigma_d/[\rho_{g0}\eta r]
\label{eq:sigmad}
\end{equation}
that may be more suitable for characterizing dust-clumping than the metallicity $Z$.
The relation with the traditional parameters can be expressed as 
$\sigma_d= (2\pi)^{1/2}Z/\Pi \approx 2.5\, Z/\Pi$.
Hence, a hypothesis is proposed where the degree of dust-clumping is determined by a set of two key parameters $(\tau_s, \sigma_d)$.



\begin{deluxetable*}{lllllllllrr}[bht!]
\tablecaption{Model parameters.}
\tablecolumns{11}
\tabletypesize{\scriptsize}
\tablehead{
\colhead{Model} &
\colhead{$\tau_s$} &
\colhead{$\sigma_d$} &
\colhead{$\Pi$} &
\colhead{$Z$} &
\colhead{$\left(\frac{L_x}{\eta r}, \frac{L_y}{\eta r}, \frac{L_z}{\eta r}\right)$} & 
\colhead{$\frac{L_z/2}{H_g}$} &
\colhead{$\frac{H_{d0}}{\eta r}$} &
\colhead{$(N_x, N_y, N_z)$} &
\colhead{$N_p$} &
\colhead{Orbits}
}
\startdata
(A) & 0.1 & 0.5 & 0.025 & 0.005 &
(4, 4, 4) & 0.05 & 0.2 & (144, 144, 144) & 4147200 & 10 \\
(B) & 0.1 & 0.5 & 0.05 & 0.01 &
(4, 4, 4) & 0.1 & 0.2 & (144, 144, 144) & 4147200 & 10 \\
(C) & 0.1 & 1.0 & 0.025 & 0.01 &
(4, 4, 4) & 0.05 & 0.2 & (144, 144, 144) & 4147200 & 10 \\
(CHd2) & 0.1 & 1.0 & 0.025 & 0.01 &
(4, 4, 4) & 0.05 & 0.4 & (144, 144, 144) & 4147200 & 10 \\
(CLz2) & 0.1 & 1.0 & 0.025 & 0.01 &
(4, 4, 8) & 0.1 & 0.2 & (144, 144, 288) & 4147200 & 10 \\
(CNp2) & 0.1 & 1.0 & 0.025 & 0.01 &
(4, 4, 4) & 0.05 & 0.2 & (144, 144, 144) & 8294400 & 10 \\
(D) & 0.1 & 1.0 & 0.05 & 0.02 & 
(4, 4, 4) & 0.1 & 0.2 & (144, 144, 144) & 4147200 &10 \\
(E) & 0.1 & 2.0 & 0.025 & 0.02 &
(4, 4, 4) & 0.05 & 0.2 & (144, 144, 144) & 4147200 & 10 \\
(F) & 0.1 & 2.0 & 0.05 & 0.04 &
(4, 4, 4) & 0.1 & 0.2 & (144, 144, 144) & 4147200 & 10 \\
(G) & 0.01 & 0.5 & 0.025 & 0.005 &
(4, 4, 4) & 0.05 & 0.2 & (288, 288, 288) & 33177600 & 50 \\
(H) & 0.01 & 0.5 & 0.05 & 0.01 &
(4, 4, 4) & 0.1 & 0.2 & (288, 288, 288) & 33177600 & 50 \\
(I) & 0.01 & 1.0 & 0.025 & 0.01 &
(4, 4, 4) & 0.05 & 0.2 & (288, 288, 288) & 33177600 & 50 \\
(J) & 0.01 & 1.0 & 0.05 & 0.02 &
(4, 4, 4) & 0.1 & 0.2 & (288, 288, 288) & 33177600 & 50 \\
(K) & 0.01 & 2.0 & 0.025 & 0.02 &
(4, 4, 4) & 0.05 & 0.2 & (288, 288, 288) & 33177600 & 50 \\
(L) & 0.01 & 2.0 & 0.05 & 0.04 &
(4, 4, 4) & 0.1 & 0.2 &(288, 288, 288) & 33177600 & 50 \\
\enddata
\end{deluxetable*}


\section{Numerical simulations and results} \label{sec:3}

Here, through three-dimensional numerical simulation, we examine whether the dust layer behaves similarly when a two-parameter set $(\tau_s, \sigma_d)$ is identical, even if the values of $Z$ and $\Pi$ are different. 
We choose values of $Z$ above and below the critical value that divides the suspension regime and streaming regime from Fig. 8 of \citet{Carrera_etal15} for $\Pi=0.05$.
We also calculate the corresponding value of $\sigma_d$ and
change the values of $Z$ and $\Pi$ by keeping $\sigma_d$ constant.
The parameters employed are listed in Table 1.

We use Athena code v. 4.2 for dust particles and gas \citep{Bai_Stone10a} and
employ a local shearing box in which $(x, y, z)$ represents the radial, azimuthal, and vertical directions. 
We also assume that the gas is isothermal,
and $c_s$ and $\Omega_K$ are constant in the shearing box. 
The vertical gravitational acceleration $-\Omega_K^2 z$ is exerted both on the gas and dust particles.
The initial gas density is assumed uniform in the $x$-- and $y$-- directions, and stratified in the $z$--direction owing to vertical gravity such that
$\rho_g=\rho_{g0} \exp[-z^2/(2H_g^2)]$. 
The gas pressure is expressed as $P=\rho_g c_s^2$. 
Hence, the radial pressure gradient is initially equal to zero: $\partial P/ \partial x=0$.  
This setting may seem to be inadequate for simulating the streaming instability because  the streaming instability is caused by the global radial pressure gradient \citep{Youdin_Goodman05}. 
In order to overcome this contradiction,
the effect of the global radial pressure gradient is included by adding an acceleration $-2\eta v_K \Omega_K$ in the radial direction to each particle 
in the Athena code (see \citet{Bai_Stone10a, Bai_Stone10b} for detailed description of the formalization).
In the Athena code, the origin of the coordinate system moves with the rotation velocity of the gas at a fiducial radius $r$, 
which is slower than the Kepler velocity by $\eta v_K$.
In this work, we employ a coordinate system in which the origin moves with the Keplerian velocity by translating $x=x'-(2/3)\eta r$, where $x'$ is the radial coordinate used in the Athena code.
We assume that all dust particles have an identical Stokes number.

The simulation region is $-L_x/2\le x \le L_x/2$, $-L_y/2\le y \le L_y/2$, and $-L_z/2\le z \le L_z/2$. The values of $L_x/(\eta r)$, $L_y/(\eta r)$, $L_z/(\eta r)$ as well as $L_z/H_g$ are listed in Table 1.
Because $L_z/H_g \ll 1$, the initial gas density is almost equal to $\rho_{g0}$ throughout the simulation region.
We employ the shearing periodic boundary condition for $x$, periodic boundary condition for $y$, and reflective boundary condition for $z$ following \citet{Bai_Stone10b}.
The mesh numbers $(N_x, N_y, N_z)$ 
and number of super-particles $N_p$ that represent the dust particles
 are listed in Table 1.

The initial unperturbed volume density of the dust particles is assumed uniform in the $x$-- and $y$-- directions, and Gaussian in the $z$--direction,
$\rho_d=\rho_{d0} \exp[-z^2/(2H_{d0}^2)]$, 
where $H_{d0}$ is the initial scale height of the dust layer. 
The unperturbed value of the column density of the dust particles is expressed as
\begin{equation}
\Sigma_{d0}=\int_{-L_z/2}^{L_z/2} \rho_d \, dz 
= \left(2 \pi \right)^{1/2} \rho_{d0} H_{d0} \mbox{ erf } [L_z/(2\sqrt{2} H_{d0})],
\label{eq:Sigmad0}
\end{equation}
where $\mbox{erf }(u) \equiv \pi^{-1/2} \int_{-u}^{u} \exp(-v^2)\, dv$.
In the following, we assume $H_{d0} \ll L_z/2$, 
such that $\mbox{ erf } [L_z/(2\sqrt{2} H_{d0})] \approx 1$.
Given a value of $\sigma_d$, the initial unperturbed value of the dust to gas density ratio at the midplane is expressed as a function of $H_{d0}/\eta r$ from Eqs. (\ref{eq:sigmad}) and (\ref{eq:Sigmad0}) as follows:
\begin{equation}
\rho_{d0}/\rho_g=\sigma_d / [(2 \pi )^{1/2} (H_{d0}/ \eta r)] \approx 0.40 \, \sigma_d (H_{d0}/ \eta r)^{-1}.
\label{eq:rhod0rhog}
\end{equation}
The value of $H_{d0}$ can be chosen arbitrarily, because the dust settling due to vertical gravity and the diffusion due to turbulence caused by the streaming and Kelvin--Helmholtz instabilities would erase the memory of the initial conditions.
In order to save computational time 
in waiting for the dust to settle, 
we employed $H_{d0}/(\eta r)=0.2$;
this value is determined to fit the results of \citet{Carrera_etal15} in the suspension regime.
We initially distribute the super-particles by employing uniform random numbers in the $x$-- and $y$--directions, and normal random number in the $z$--direction for determining the initial positions of the super-particles.
Hence, the seeds of the streaming instability are the initial perturbation of the dust volume density caused by the random distribution of the super-particles.
Note that the results are almost similar even if the initial values of $H_{d0}/\eta r$, $\rho_{d0}/\rho_g$, $L_z/(\eta r)$ and $N_p$ are different as long as the sets of values $(\tau_s, \sigma_d)$ are identical,
as shown in Appendix A.

The semi-implicit method for the time-integration of the particles and gas is adopted.
We determine the Courant number such that the numerical instability due to the back reaction of the dust on gas is avoided: 
$\Delta t < 0.5 (\rho_g/\rho_{d \mbox{max} }) t_{stop} $,
where $\rho_{d\mbox{max}}$ is the maximum dust density.
If the Courant number determined from the above condition is greater than 0.4, 
we employ 0.4 in order to satisfy the usual Courant condition.

Figure 1 shows the column density of dust $\Sigma_d$ normalized by the unperturbed value $\Sigma_{d0}$ in the $x$--$y$ plane at $t \Omega_K/(2\pi)=10$ for models (A)--(F) with $\tau_s=0.1$.
The red color indicates that $\Sigma_d > \Sigma_{d0}$, 
and the blue color indicates that $\Sigma_d < \Sigma_{d0}$.
Figure 2 shows the mean volume density of dust  
$\left< \rho_d \right>_y=\sum_{i_y=1}^{N_y} \rho_{d, i_y} /N_y$,
where $i_y$ represents the mesh number in the $y$ direction, 
normalized by the unperturbed gas density at the midplane $\rho_{g0}$
in the $x$--$z$ plane at $t \Omega_K/(2\pi)=10$ for models (A)--(F).
The red color indicates that $\left< \rho_d \right>_y > \rho_{g0}$ 
and the blue color indicates that $\left< \rho_d \right>_y  < \rho_{g0}$.

Radial clumps of particles are observed in model (C) at approximately $x/\eta r=-0.4$, 0.6, and 1.5 (including a weak clump at approximately $x/\eta r=-1.5$).
In contrast, in models (A) and (B), the clumping is weak, and the clumps are not 
radially confined.
The density distribution of particles in model (B) is similar to that in model (A) and is very different from that in model (C);
however, the value of $Z$ for model (B) is same as that for model (C) and is twice as large as that for model (A).
In contrast, the value of $\sigma_d$ for model (B) is identical to that for model (A).
Model (D) also shows radial clumping at approximately $x/\eta r=-1.5$ and 1.1.
The magnitude of radial-clumping in model (D) is similar to that in model (C),
and the values of $\sigma_d$ for models (C) and (D) are identical.
The clumping of particles develops more extensively in models (E) and (F),
and the values of $\sigma_d$ for models (E) and (F) are identical, 
although the 
value
of $Z$ for model (E) is a half the value of $Z$ for model (F).
Hence, a new parameter $\sigma_d$ defined by Eq. (\ref{eq:sigmad}) is more appropriate than $Z$ for characterizing the particle clumping.

Figure 3 shows the maximum dust density $\rho_{d\, \mbox{max}}$ normalized by the unperturbed gas density at the midplane as a function of time.
The asymptotic value of  $\rho_{d\, \mbox{max}} / \rho_{g0}$ is about 10 for models (A) and (B) and 30--40 for models (C) and (D). 
In contrast, in models (E) and (F),
$\rho_{d\, \mbox{max}} / \rho_{g0}$ continues to grow 
toward the critical density of the gravitational instability 
normalized by the gas density,
which is a few hundred \citep{Yamoto_Sekiya04}.
Note that our simulations do not take self-gravity into account, and the increase in the dust volume density is caused by the streaming instability.
In order to elucidate the final state of models (E) and (F), numerical simulations that include self-gravity are necessary 
as have been performed by \citet{Simon_etal16, Simon_etal17}.
In any case, the above results also support the hypothesis 
that $\sigma_d$ is appropriate for characterizing the tendency 
for particle clumping for a fixed value of $\tau_s$.

Figures 4 and 5 show $\Sigma_d / \Sigma_{d0}$ in the $x$--$y$ plane and $\left< \rho_d \right>_y / \rho_{g0}$ in the $x$--$z$ plane, 
respectively,
at $t \Omega_K/(2\pi)=50$ for models (G)--(L) with $\tau_s=0.01$. 
In models (G) and (H), it can be seen that radial clumping is weak and faint.
The particle distribution in model (H) is similar to that in model (G),
and the values of $\sigma_d$ for models (G) and (H) are identical.
In contrast, the value of $Z$ for model (H) is same as that for model (I)
and is twice as large as that of model (G).
Moreover, several clear radial clumps are 
also observed
in models (I)  and (J), 
and the values of $\sigma_d$ for models (I) and (J) are identical.
In model (K), a very dense isolated clump can be observed at approximately $x/\eta r=-1.6$, and a broad radial clump can be observed at approximately $1.5 \la x/\eta r \la 2$.
Similarly, in model (L), an isolated clump and a broad radial clump can be observed at approximately $x/\eta r =\pm2.0$  at about $0.9 \la x/\eta r \la 1.4$, respectively.
The formation and evolution of the isolated clump and broad radial clump in models (K) and (L) proceed very similarly, as seen in the animations of Figs. 4 (K) and (L),
and the values of $\sigma_d$ for models (K) and (L) are identical.

Figure 6 presents the time evolution of the maximum dust density for 
models (G)--(L).
The maximum dust density converges at approximately 2.4 in models (G) and (H) with $\sigma_d=0.5$, and approximately 3.8 in models (I) and (J) with $\sigma_d=1.0$.
In contrast, the maximum density continues to grow to reach about 50 in models (K) and (L) with $\sigma_d=2.0$.
The results for $\tau_s=0.01$ also indicate that $\sigma_d$ is more appropriate than $Z$ for characterizing the particle clumping.

Hence, the hypothesis that 
particle clumping
is classified  by a set of two parameters $(\tau_s, \sigma_d)$
is confirmed.

\section{Conclusions and Discussion} \label{sec:4}

In previous studies, 
dust clumping caused by the streaming instability
was classified by employing a set of three dimensionless parameters
$(\tau_s, Z, \Pi)$.
A new parameter $\sigma_d= (2\pi)^{1/2}Z/\Pi$ was introduced 
in order to simplify
the classification, 
where $\sigma_d$ is the dust column density $\Sigma_d$
nondimensionalized using the unperturbed midplane gas density, the Keplerian angular velocity, and the difference 
between
the Keplerian velocity and gas velocity in the region outside 
the dust layer 
where the dust 
volume
density is much less than the 
gas
density.
The results of the numerical simulations indicate that the extent of 
particle clumping 
is similar for an identical set of $(\tau_s, \sigma_d)$, even if the values of $Z$ and $\Pi$ are different.
Hence, a set of two parameters $(\tau_s, \sigma_d)$ is sufficient to classify the extent of the particle clumping
 caused by the streaming instability.

\citet{Carrera_etal15} performed numerical simulations of the dust layer by assuming $\Pi=0.05$ for obtaining Fig. 8, 
which was revised by \citet{Yang_etal17} as their Fig. 9.
If a set of parameters $(\tau_s, Z, \Pi)$ is given, these figures
can be used to determine the behavior of the dust layer 
for a different value of $\Pi$.
The corresponding value of the metallicity for $\Pi=0.05$  with 
the
same value of $\sigma_d$ is expressed as
$Z|_{\Pi=0.05}=0.05(Z/\Pi)$.
Applying this value of the metallicity with $\tau_s$ to 
Fig. 8 of \citet{Carrera_etal15} and 
Fig. 9 of \citet{Yang_etal17},
we can determine whether the dust layer is in the suspension region or the streaming region.

The results of numerical simulation in this work show that 
the dust-clumping grows if $\sigma_d \ga 1$.
Given that $\sigma_d$ is a measure of dust-to-gas mass-ratio in the dust layer, as explained in section 2,
the results indicate that the dust-clumping grows if the inertia of the dust particles is greater than that of the gas.

In this work, we performed numerical simulations for $\tau_s=0.1$ and 0.01.
At lower values of $\tau_s$, 
the evolution time will be significantly long because of the low relative velocities between the dust particles and gas.
In that regard, new formulations for dust and gas by \citet{Laibe_Price14} and \citet{Lin_Youdin17} may be useful for resolving this issue, 
which will also be addressed in future work.

\acknowledgments

This work was supported by JSPS KAKENHI Grant Number JP15K05268.
Numerical computations were carried out on the Cray XC30 at the Center for Computational Astrophysics, National Astronomical Observatory of Japan.

%




\appendix
\section{
Dependence of maximum dust volume density on initial scale height of the dust layer, vertical size of the computational region, and number of super particles
}

Here, we examine whether the asymptotic value of $\rho_{d,\mbox{max}}/\rho_{g0}$ depends on the initial scale height of the dust layer $H_{d0}/(\eta r)$, the vertical size of the simulation box $L_z/(\eta r)$, and the number of super particles $N_p$ for an identical set of values of $\tau_s$ and $\sigma_d$.
We employ model (C) as a fiducial model.
As shown in Table 1, model (CHd2) has twice the value of $H_{d0}/(\eta r)$, 
model (CLz2) has twice the values of $L_z/(\eta r)$ and $N_z$, 
and model (CNp2) has twice the value of $N_p$, when compared with model (C).
We also show the results of model (A) that has half value of $\sigma_d$ and the identical value of $\tau_s$ when compared with model (C).

Figure 7 (a) and (b) show the time evolution of the dust density at the midplane averaged over all the $x$-- and $y$--meshes, $\left< \left. \rho_d \right|_{z=0} \right>_{xy}$, and
the maximum dust density among all the meshes, $\rho_{d,\mbox{max}}$, normalized by $\rho_{g0}$.
The initial value of $\left< \left. \rho_d \right|_{z=0} \right>_{xy} / \rho_{g0}$ is exactly equal to $\rho_{d0}/\rho_g$ given by Eq. (\ref{eq:rhod0rhog}), namely, 2.0 for models (C), (CLz2) and (CNp2), and 1.0 for models (A) and (CHd2).
The initial values of $\rho_{d\mbox{max}} /\rho_{g0}$ are larger than $\rho_{d0}/\rho_{g0}$, because we employed random numbers to distribute the super particles and the number of super particles in each mesh in the neighborhood of the midplane fluctuates around the unperturbed value.

The midplane dust density increases owing to the dust settling during first 2.5 orbits 
for models (A), (C), (CLz2), and (CNp2) with $H_{d0}/(\eta r)=0.2$ and 3.5 orbits for model (CHz2) with $H_{d0}=0.4$. After the dust-settling period, $\left< \left. \rho_d \right|_{z=0} \right>_{xy}$ decreases due to diffusion of particles by the turbulence caused by the streaming and Kelvin--Helmholtz instabilities. In contrast, $\rho_{d\mbox{max}} /\rho_{g0}$ continues to grow because of the growth of the particle clumping caused by the streaming instability.
Finally, $\left< \left. \rho_d \right|_{z=0} \right>_{xy}$ and $\rho_{d\mbox{max}} /\rho_{g0}$  approaches to nearly constant values after 4 to 5 orbits. In this stage, the particle clumping and diffusion by the streaming and Kelvin--Helmholtz instabilities and the settling of dust particles due to vertical gravity are in quasi-equilibrium.

The asymptotic values of $\rho_{d\mbox{max}} /\rho_{g0}$ are similar for models (C), (CHd2), (CLz2), and (CNp2) with $(\tau_s, \sigma_d)=(0.1, 1.0)$ although we use various values for $H_{d0}/(\eta r)$, $\rho_{d0}/\rho_{g0}$, $L_z/(\eta r)$ and $N_p$. 
In contrast, the asymptotic value for model (A) with $(\tau_s, \sigma_d)=(0.1, 0.5)$
is different from that for models (C), (CHd2), (CLz2) and (CNp2) with $(\tau_s, \sigma_d)=(0.1, 1.0)$.
Hence, the maximum dust density is independent of the values of $H_{d0}/(\eta r)$, $\rho_{d0}/\rho_g$, $L_z/(\eta r)$ and $N_p$, if we keep a set of values of the two parameters $(\tau_s, \sigma_d)$ constant.

\begin{figure*}
\gridline{\fig{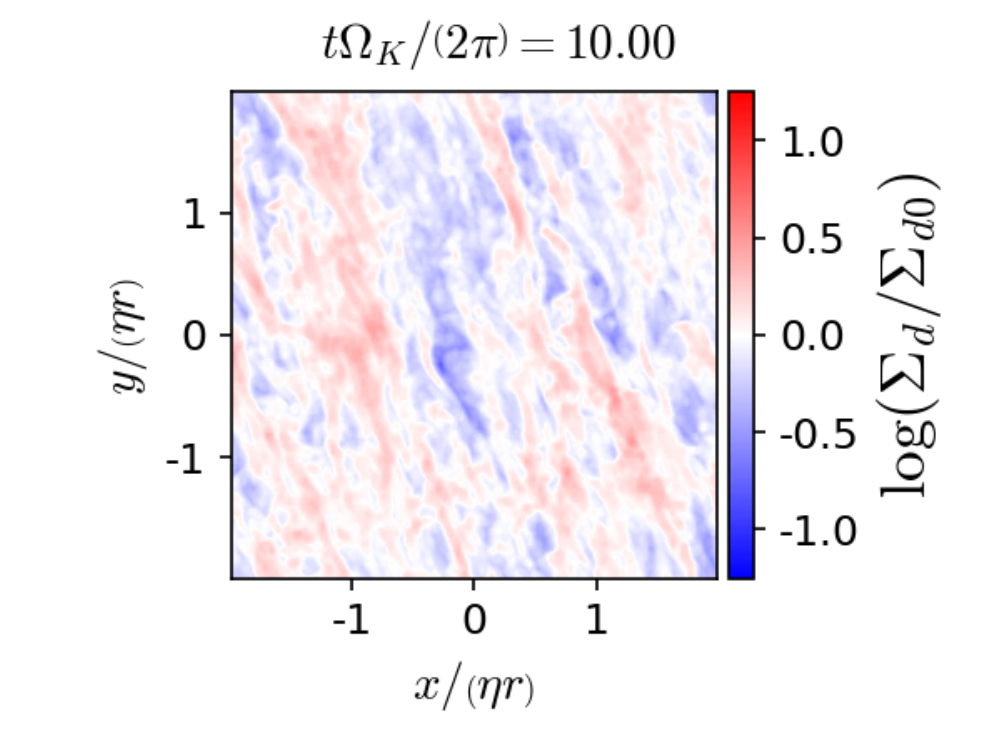}{0.45\textwidth}{(A) $\tau_s=0.1$, $\sigma_d=0.5$, $\Pi=0.025$, $Z=0.005$}
            \fig{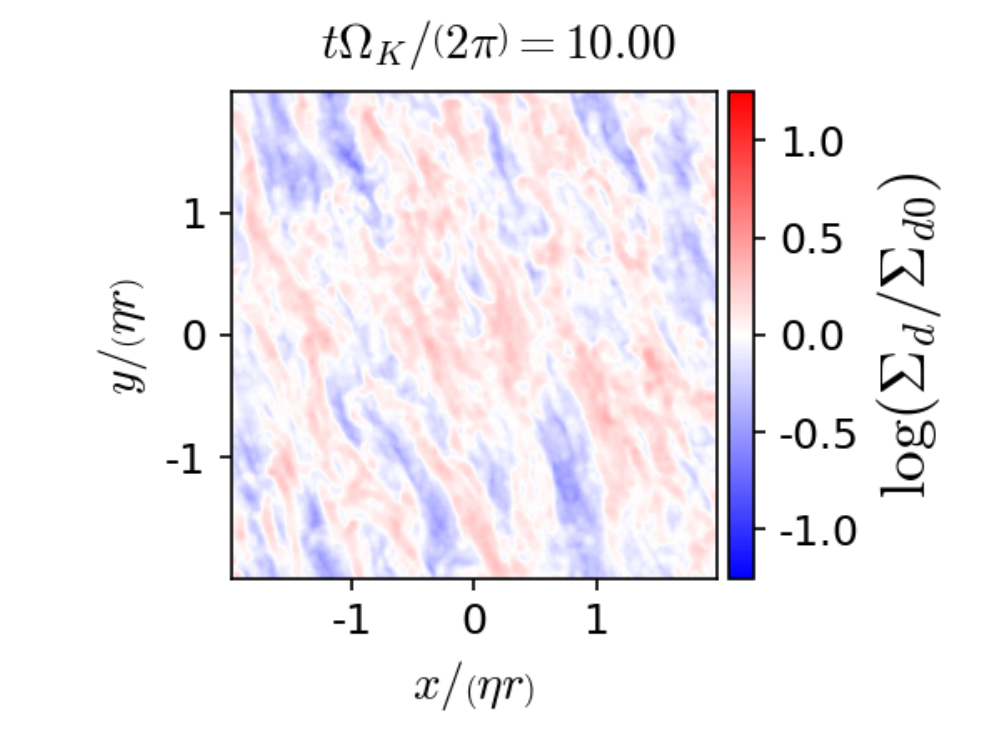}{0.45\textwidth}{(B) $\tau_s=0.1$, $\sigma_d=0.5$, $\Pi=0.05$, $Z=0.01$}
            }
\gridline{\fig{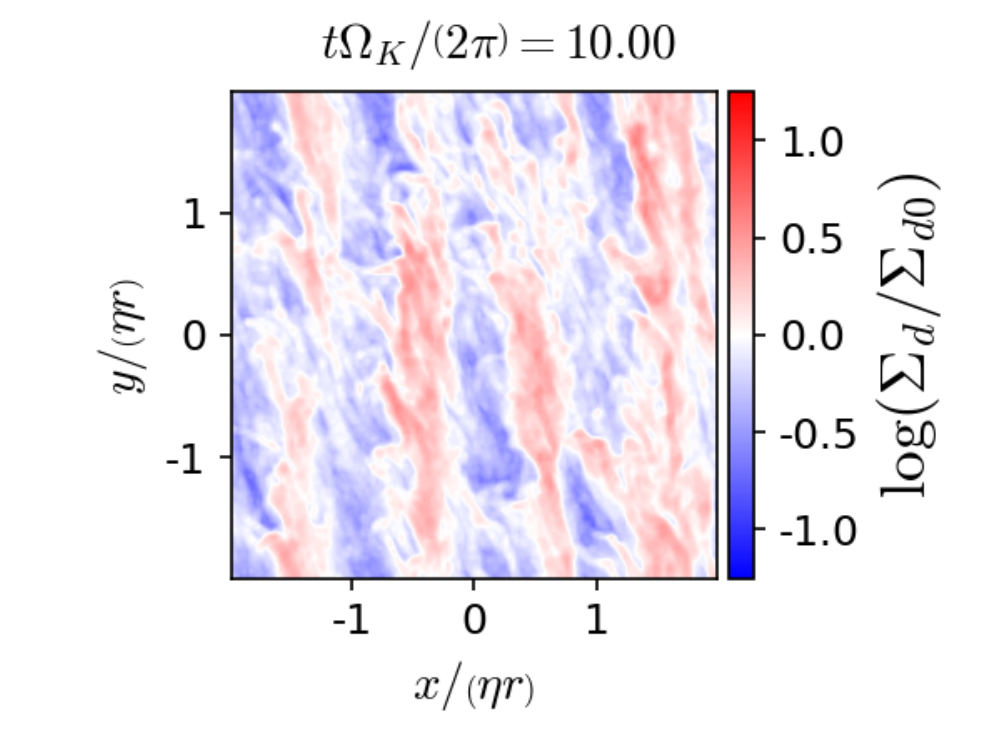}{0.45\textwidth}{(C) $\tau_s=0.1$, $\sigma_d=1.0$, $\Pi=0.025$, $Z=0.01$}
            \fig{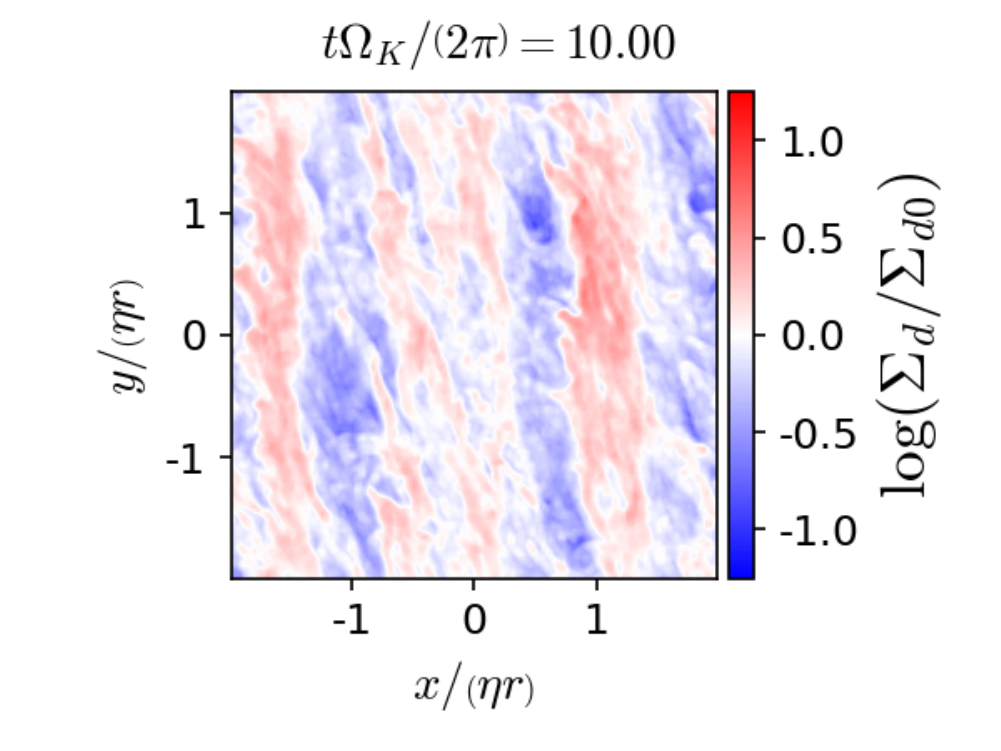}{0.45\textwidth}{(D) $\tau_s=0.1$, $\sigma_d=1.0$, $\Pi=0.05$, $Z=0.02$}
            }
\gridline{\fig{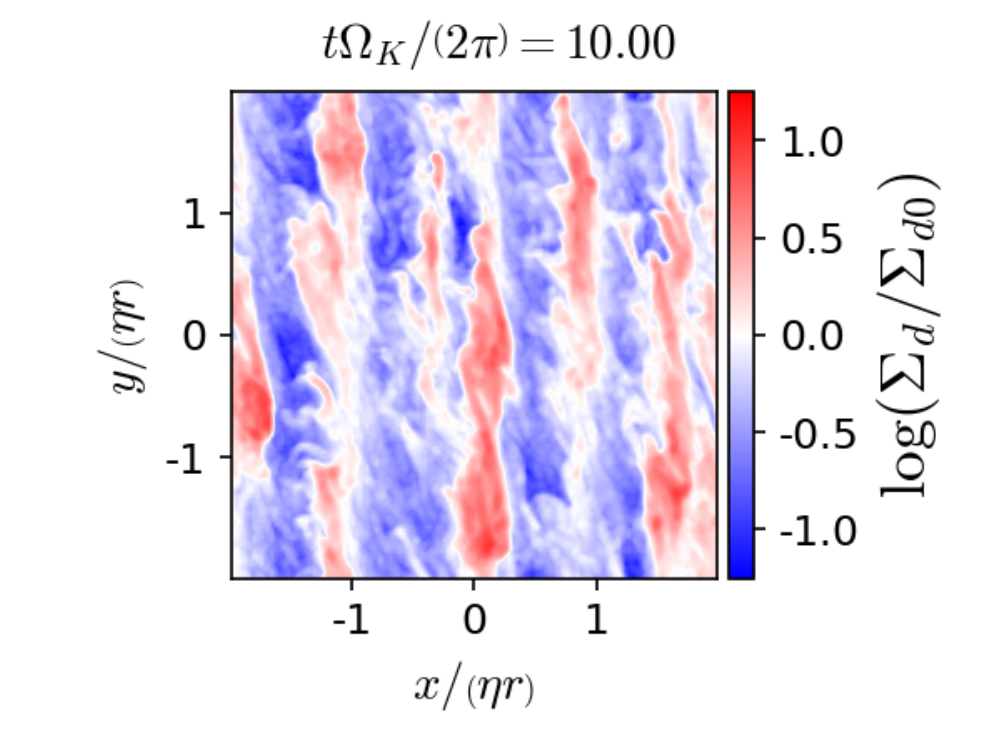}{0.45\textwidth}{(E) $\tau_s=0.1$, $\sigma_d=2.0$, $\Pi=0.025$, $Z=0.02$}
            \fig{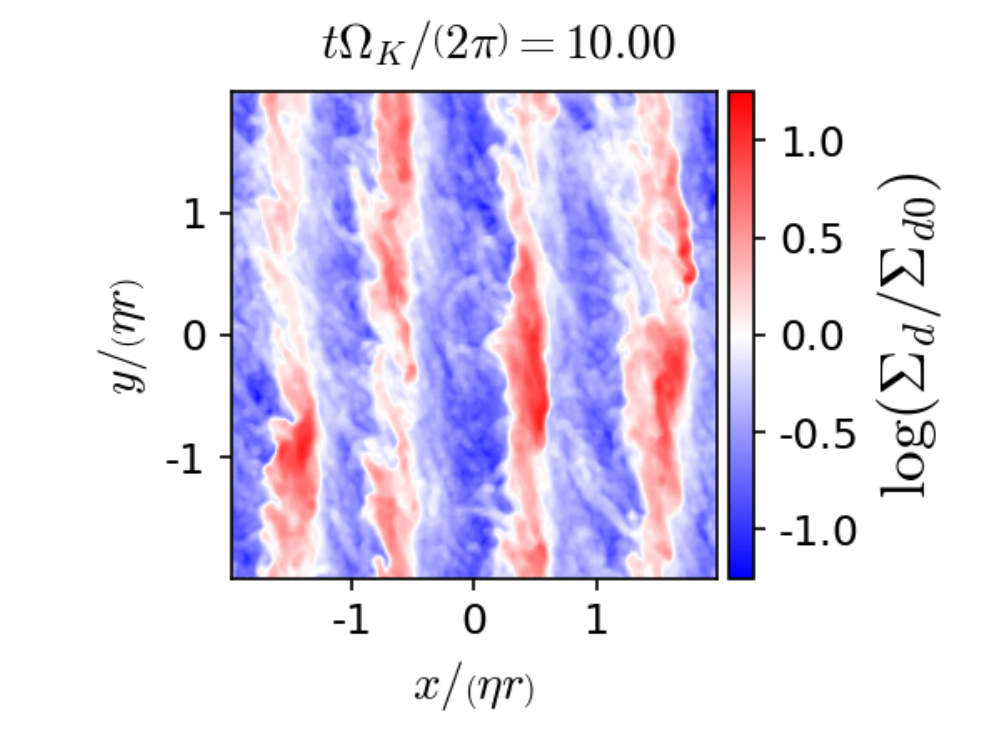}{0.45\textwidth}{(F) $\tau_s=0.1$, $\sigma_d=2.0$, $\Pi=0.05$, $Z=0.04$}
            }
\caption{Dust column density distribution at $t \Omega_K / \left( 2 \pi \right) = 10$ for models (A) to (F). 
}
\end{figure*}

\begin{figure*}
\gridline{\fig{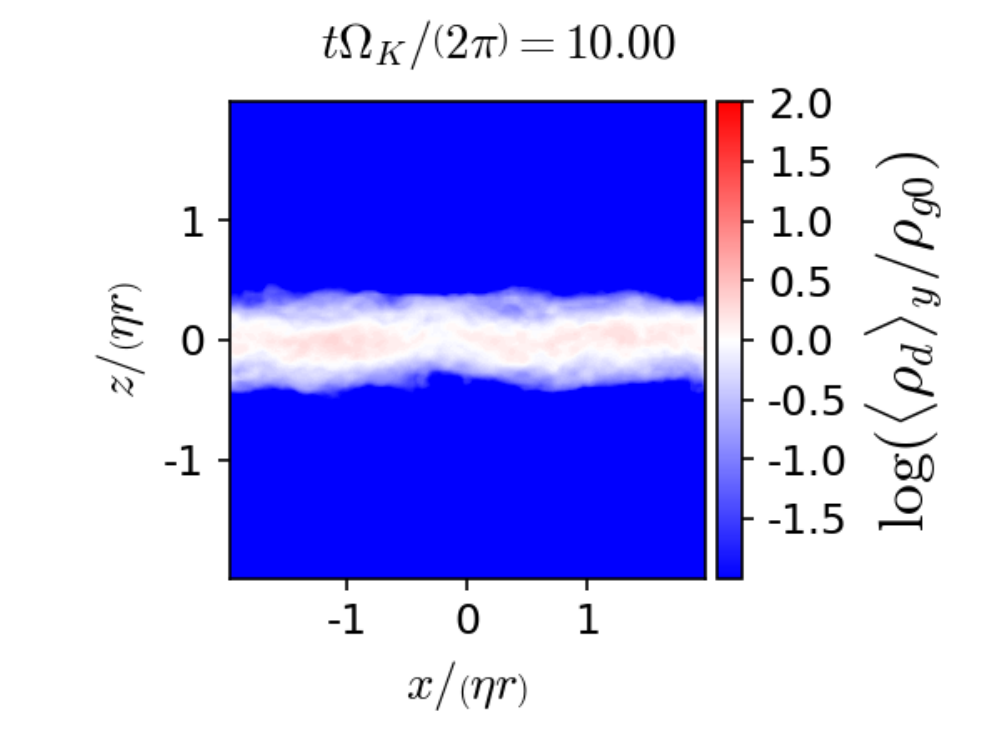}{0.45\textwidth}{(A) $\tau_s=0.1$, $\sigma_d=0.5$, $\Pi=0.025$, $Z=0.005$}
            \fig{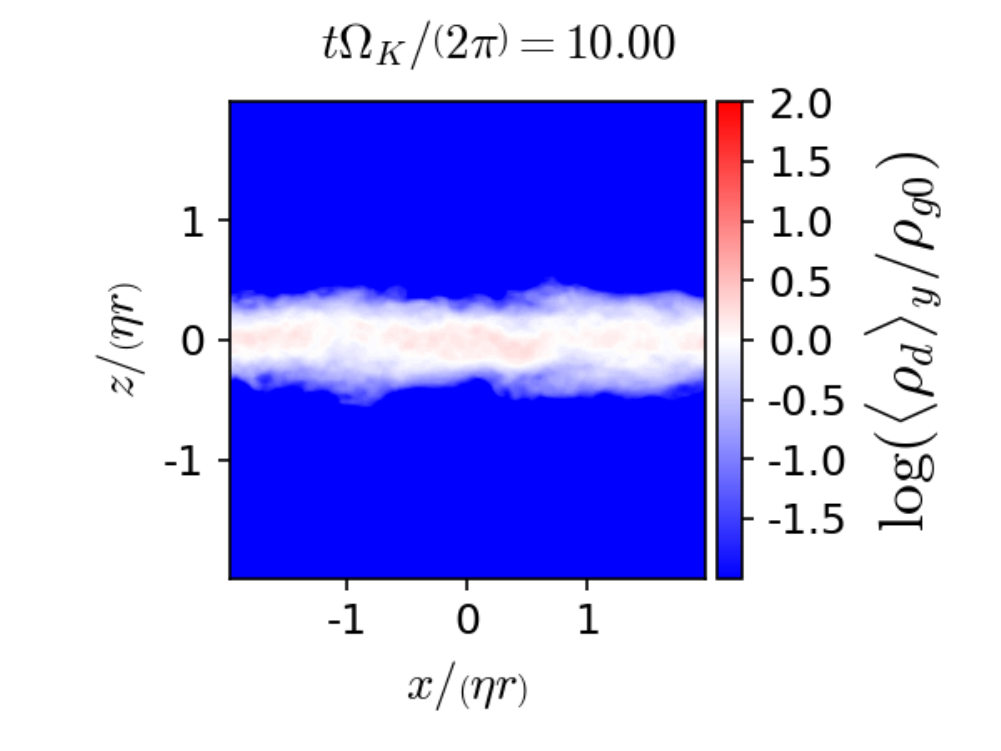}{0.45\textwidth}{(B) $\tau_s=0.1$, $\sigma_d=0.5$, $\Pi=0.05$, $Z=0.01$}
            }
\gridline{\fig{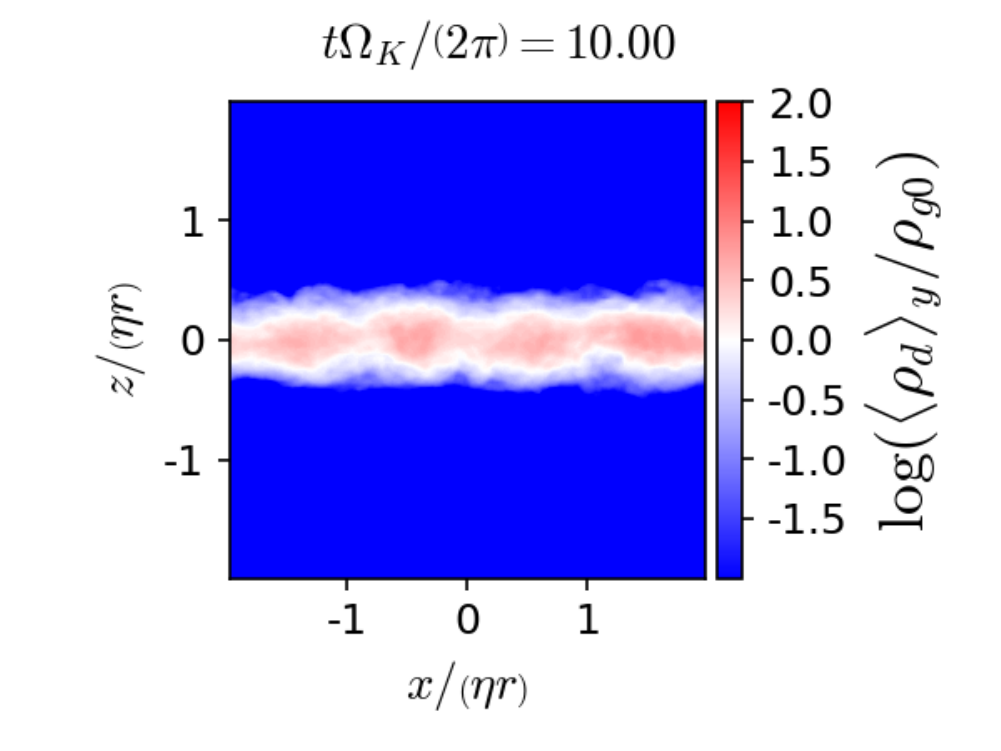}{0.45\textwidth}{(C) $\tau_s=0.1$, $\sigma_d=1.0$, $\Pi=0.025$, $Z=0.01$}
            \fig{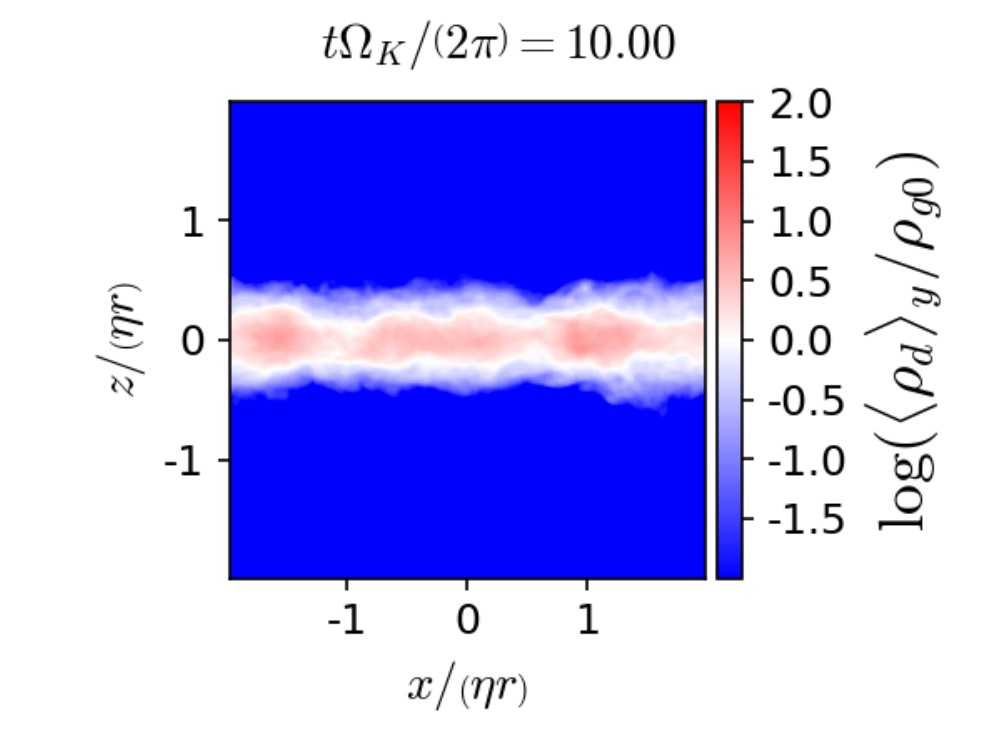}{0.45\textwidth}{(D) $\tau_s=0.1$, $\sigma_d=1.0$, $\Pi=0.05$, $Z=0.02$}
            }
\gridline{\fig{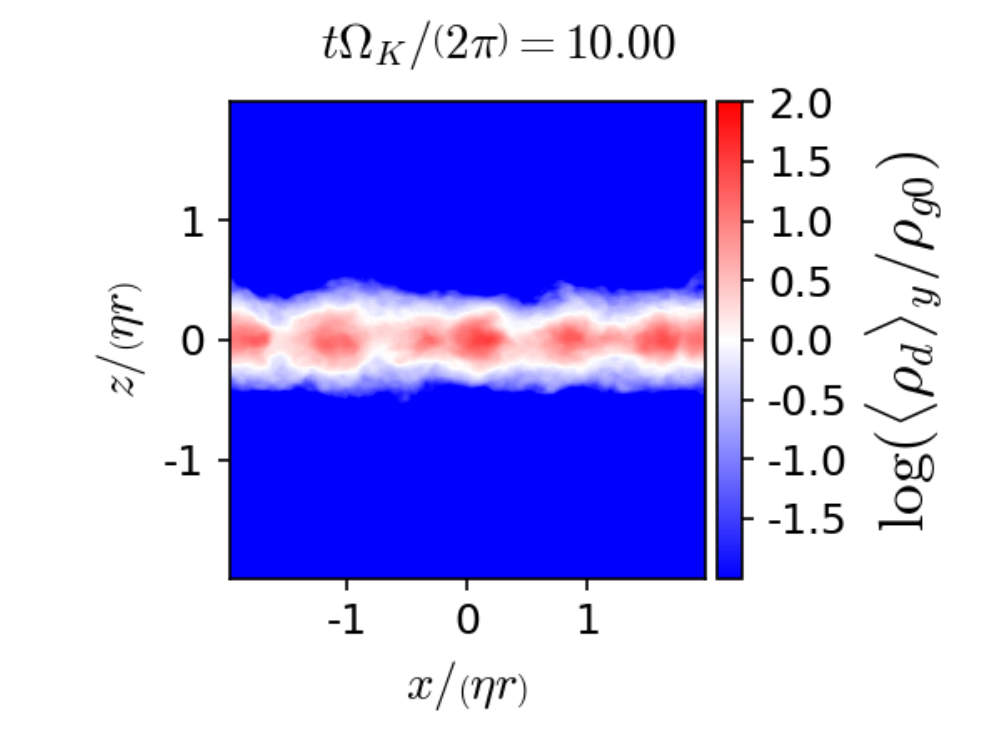}{0.45\textwidth}{(E) $\tau_s=0.1$, $\sigma_d=2.0$, $\Pi=0.025$, $Z=0.02$}
            \fig{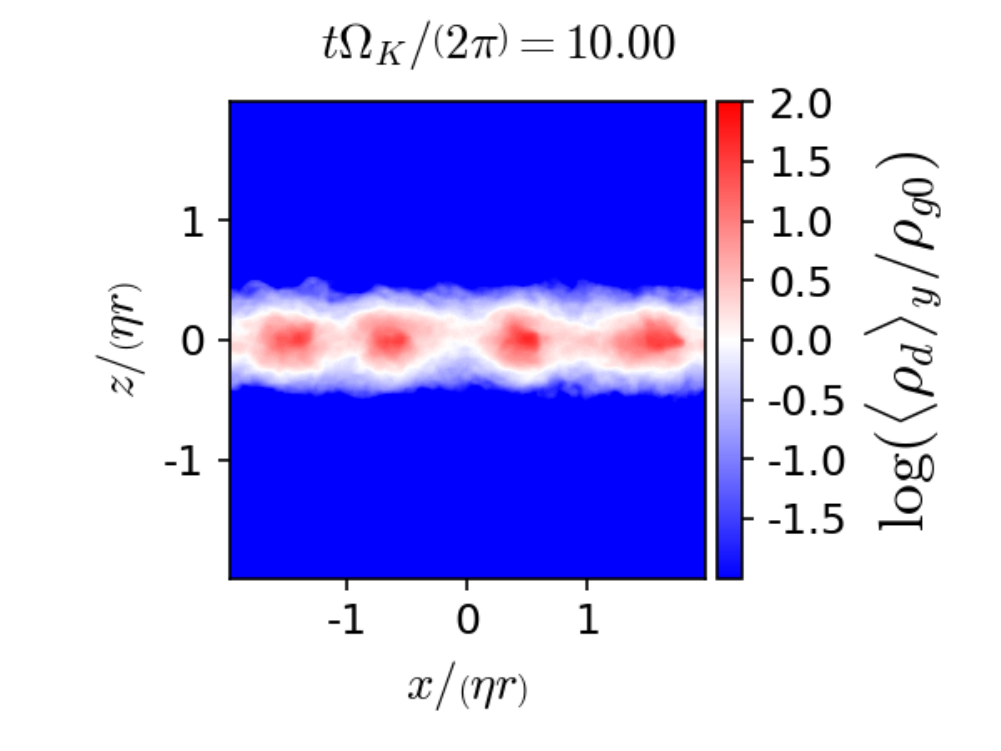}{0.45\textwidth}{(F) $\tau_s=0.1$, $\sigma_d=2.0$, $\Pi=0.05$, $Z=0.04$}
            }
\caption{$y$-averaged dust density as a function of $(x, z)$
at $t \Omega_K / \left( 2 \pi \right) = 10$
for models (A) to (F).  
}
\end{figure*}

\begin{figure*}
\gridline{\fig{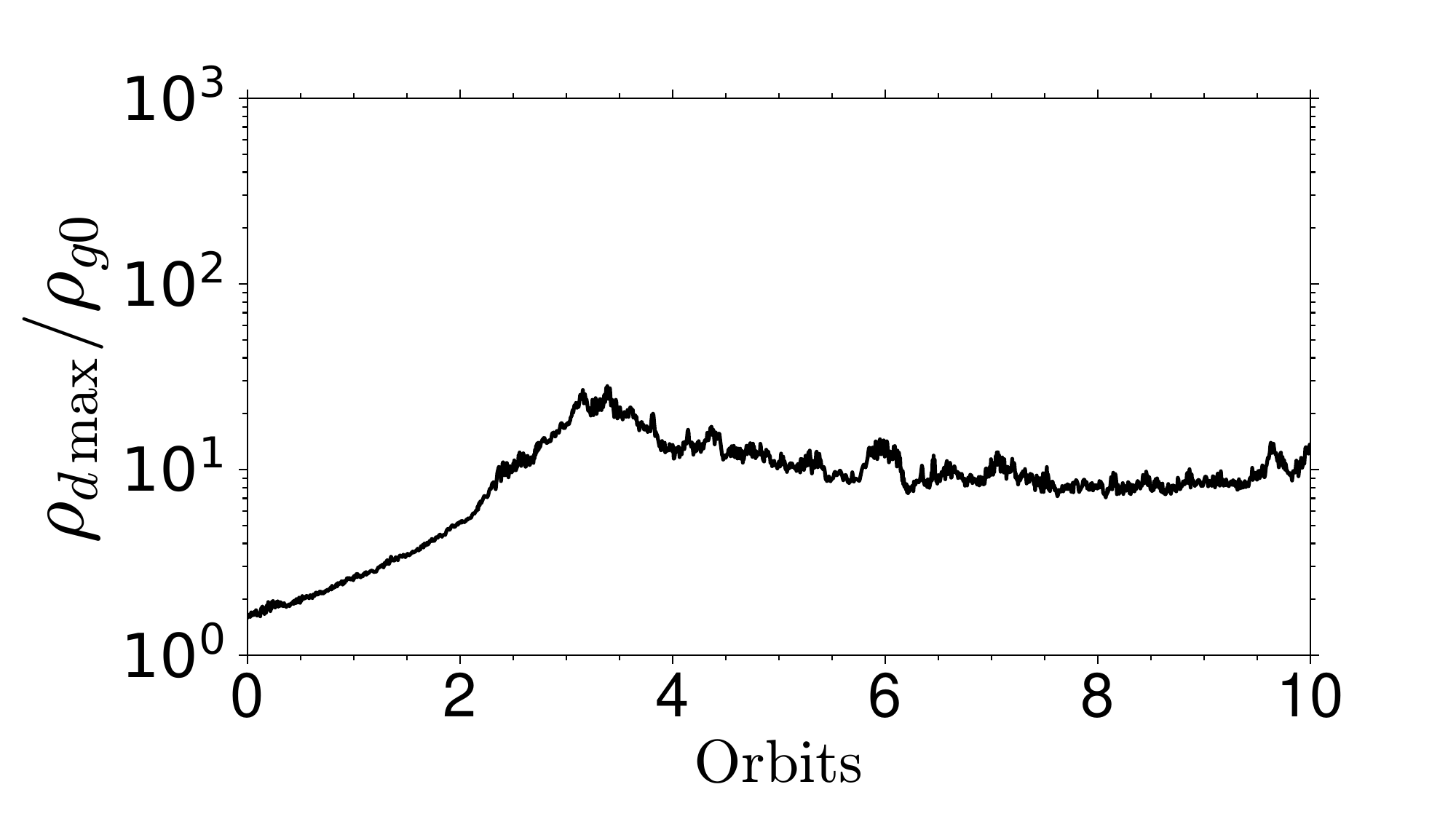}{0.45\textwidth}{(A) $\tau_s=0.1$, $\sigma_d=0.5$, $\Pi=0.025$, $Z=0.005$}
            \fig{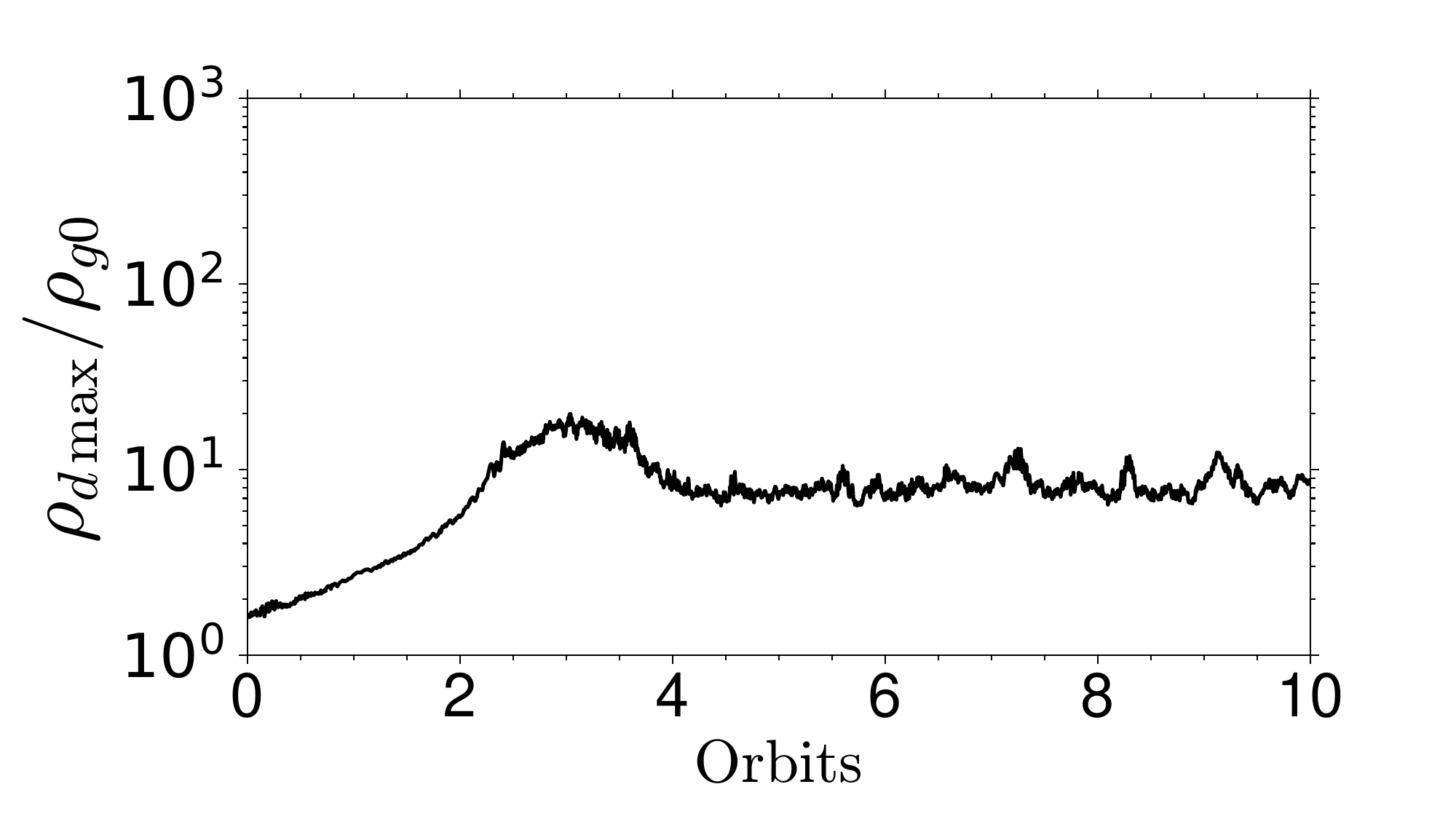}{0.45\textwidth}{(B) $\tau_s=0.1$, $\sigma_d=0.5$, $\Pi=0.05$, $Z=0.01$}
            }
\gridline{\fig{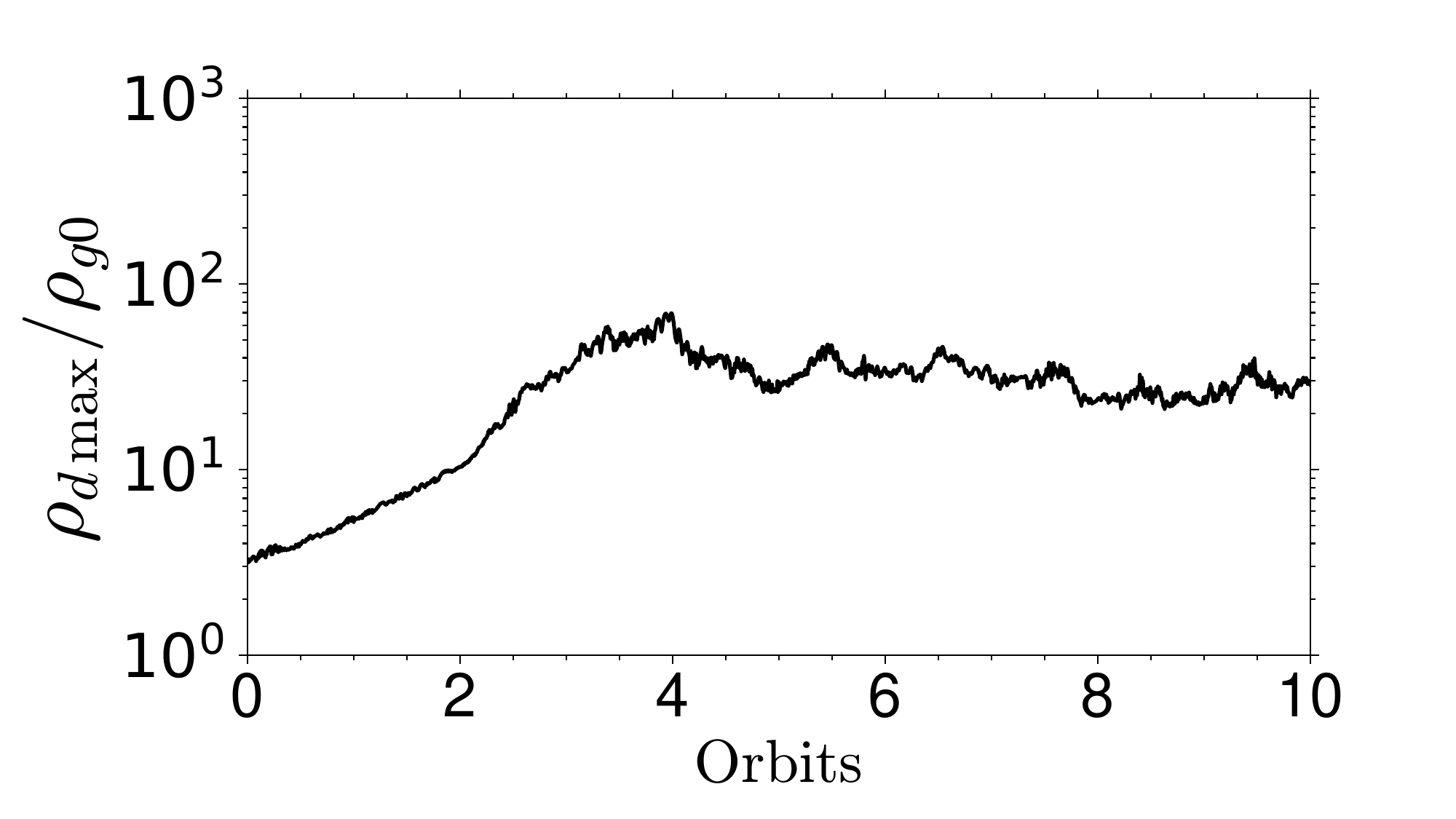}{0.45\textwidth}{(C) $\tau_s=0.1$, $\sigma_d=1.0$, $\Pi=0.025$, $Z=0.01$}
            \fig{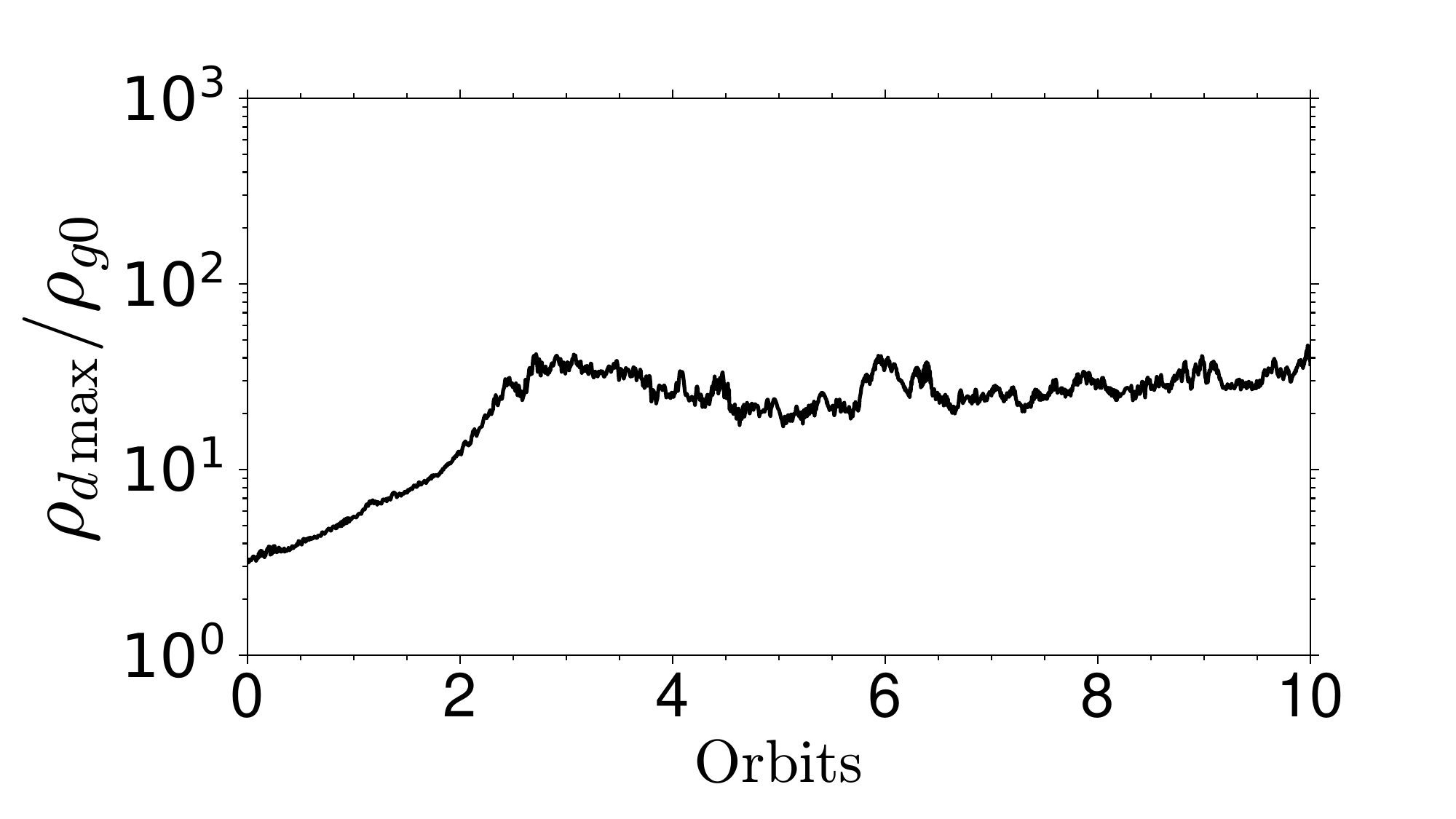}{0.45\textwidth}{(D) $\tau_s=0.1$, $\sigma_d=1.0$, $\Pi=0.05$, $Z=0.02$}
            }
\gridline{\fig{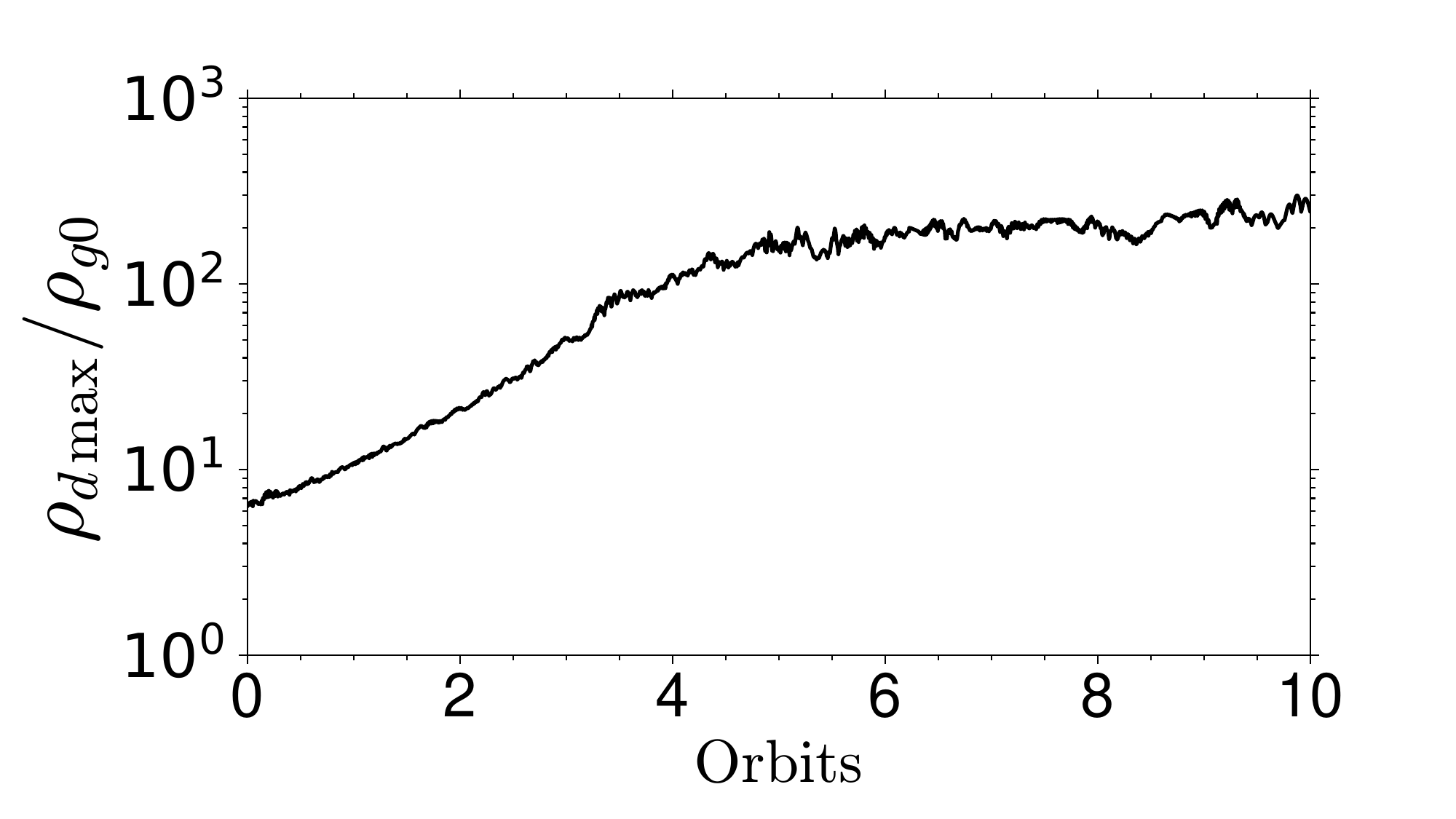}{0.45\textwidth}{(E) $\tau_s=0.1$, $\sigma_d=2.0$, $\Pi=0.025$, $Z=0.02$}
            \fig{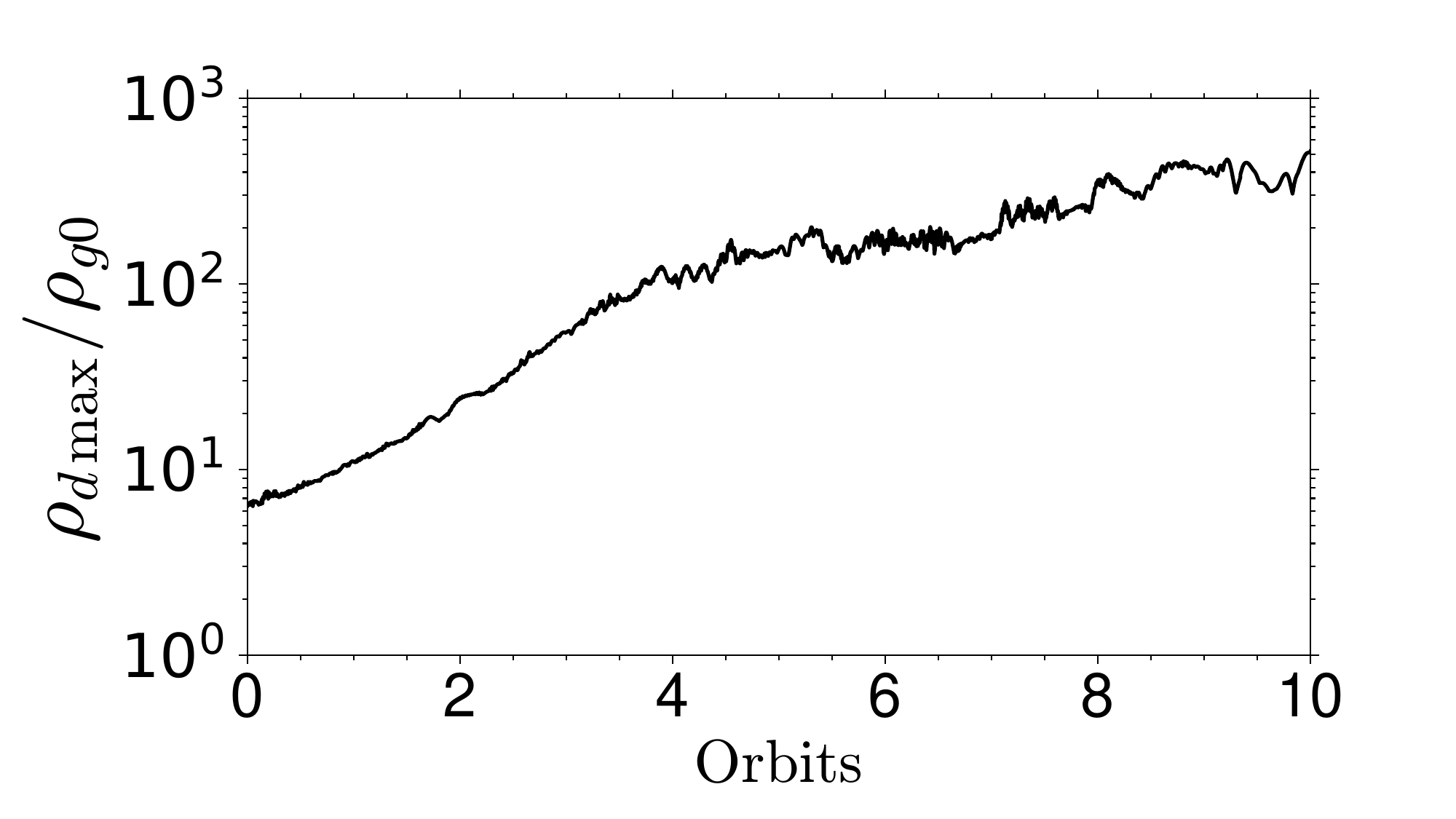}{0.45\textwidth}{(F) $\tau_s=0.1$, $\sigma_d=2.0$, $\Pi=0.05$, $Z=0.04$}
            }
\caption{Maximum dust density as a function of time (orbits) for models (A) to (F).}
\end{figure*}

\begin{figure*}
\gridline{\fig{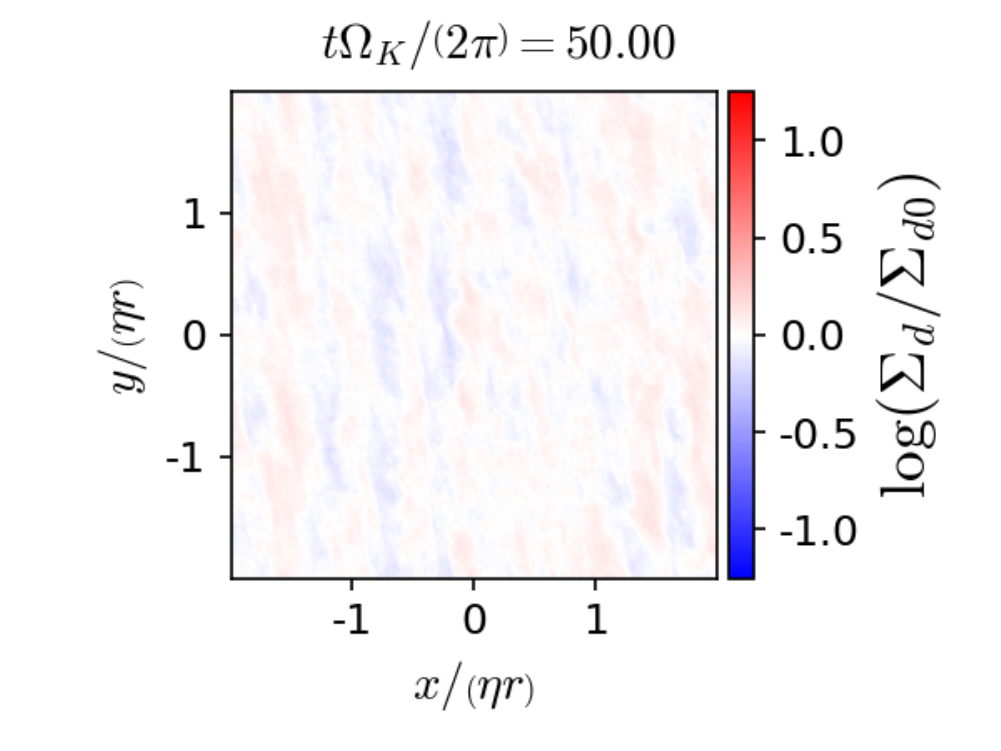}{0.45\textwidth}{(G) $\tau_s=0.01$, $\sigma_d=0.5$, $\Pi=0.025$, $Z=0.005$}
            \fig{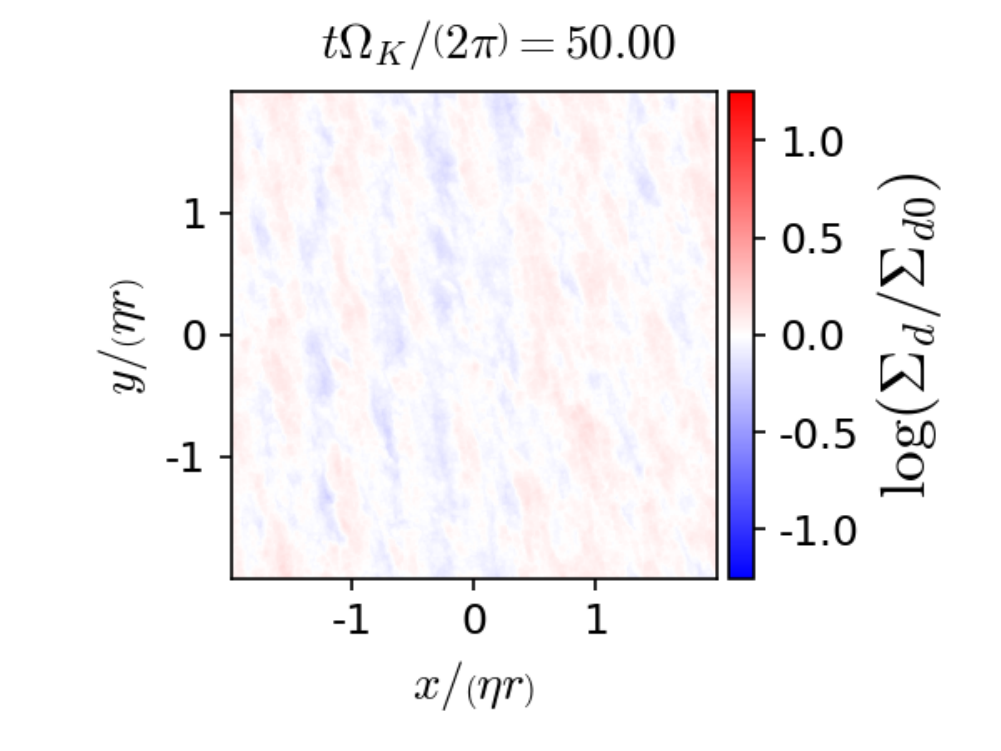}{0.45\textwidth}{(H) $\tau_s=0.01$, $\sigma_d=0.5$, $\Pi=0.05$, $Z=0.01$}
            }
\gridline{\fig{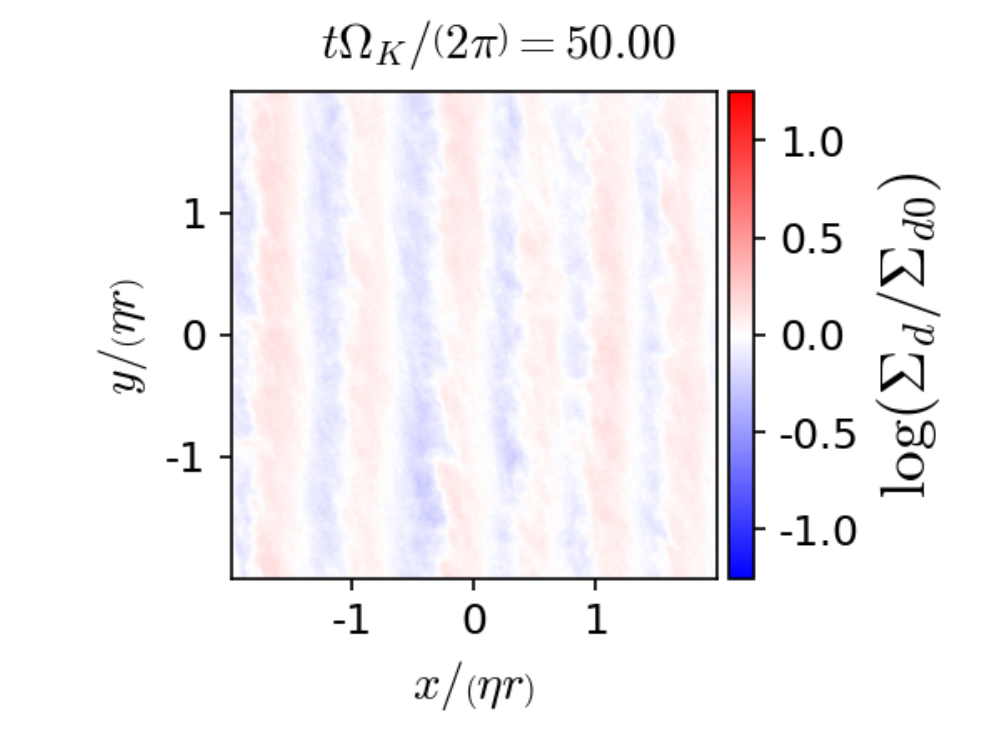}{0.45\textwidth}{(I) $\tau_s=0.01$, $\sigma_d=1.0$, $\Pi=0.025$, $Z=0.01$}
            \fig{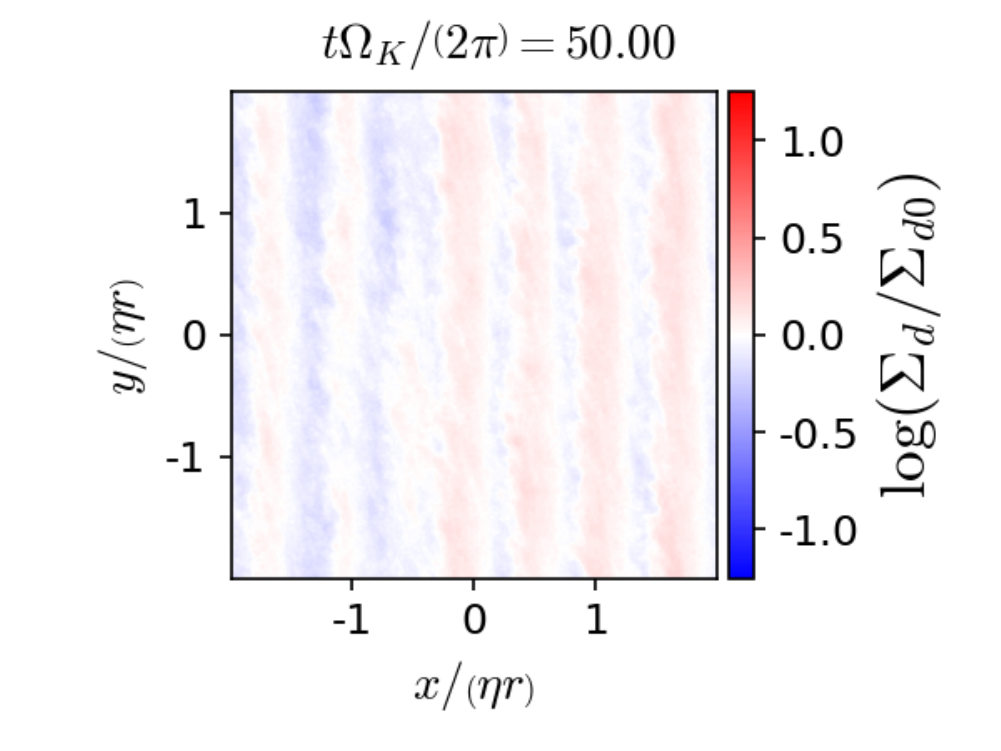}{0.45\textwidth}{(J) $\tau_s=0.01$, $\sigma_d=1.0$, $\Pi=0.05$, $Z=0.02$}
            }
\gridline{\fig{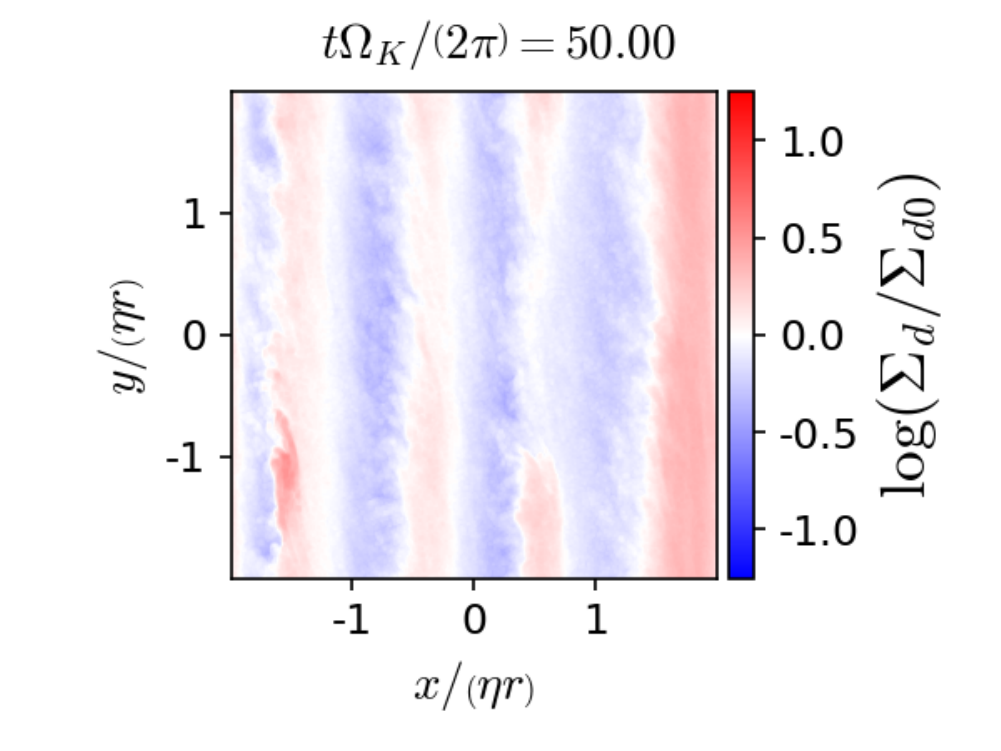}{0.45\textwidth}{(K) $\tau_s=0.01$, $\sigma_d=2.0$, $\Pi=0.025$, $Z=0.02$}
            \fig{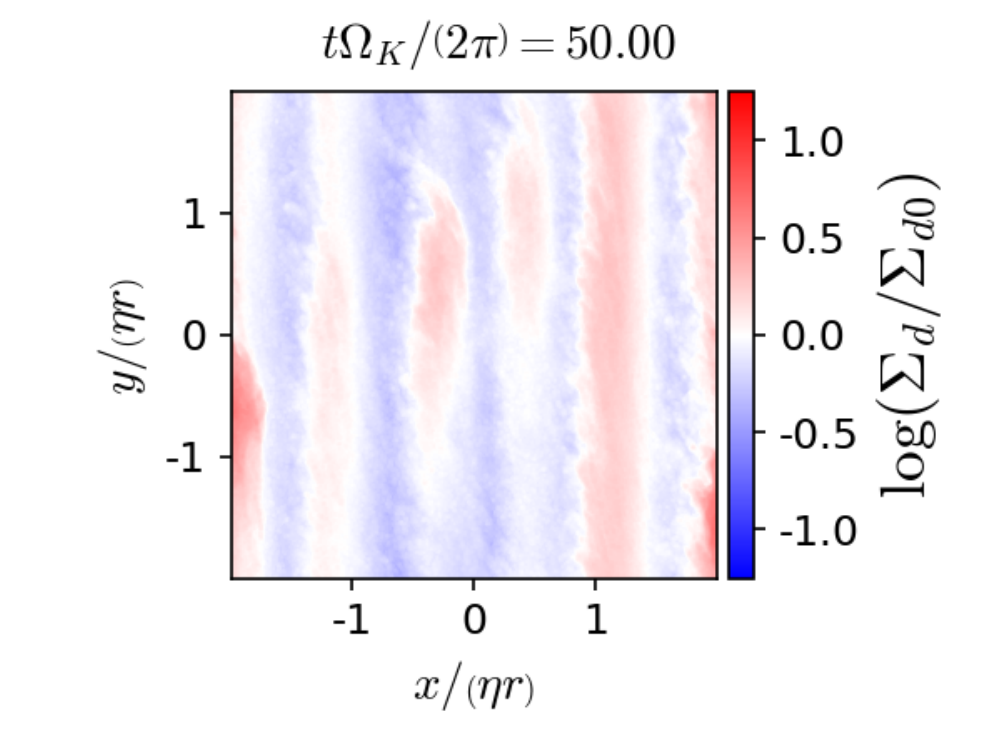}{0.45\textwidth}{(L) $\tau_s=0.01$, $\sigma_d=2.0$, $\Pi=0.05$, $Z=0.04$}
            }
\caption{
Dust column density distribution at $t \Omega_K / \left( 2 \pi \right) = 50$ for models (G) to (L).  
}
\end{figure*}

\begin{figure*}
\gridline{\fig{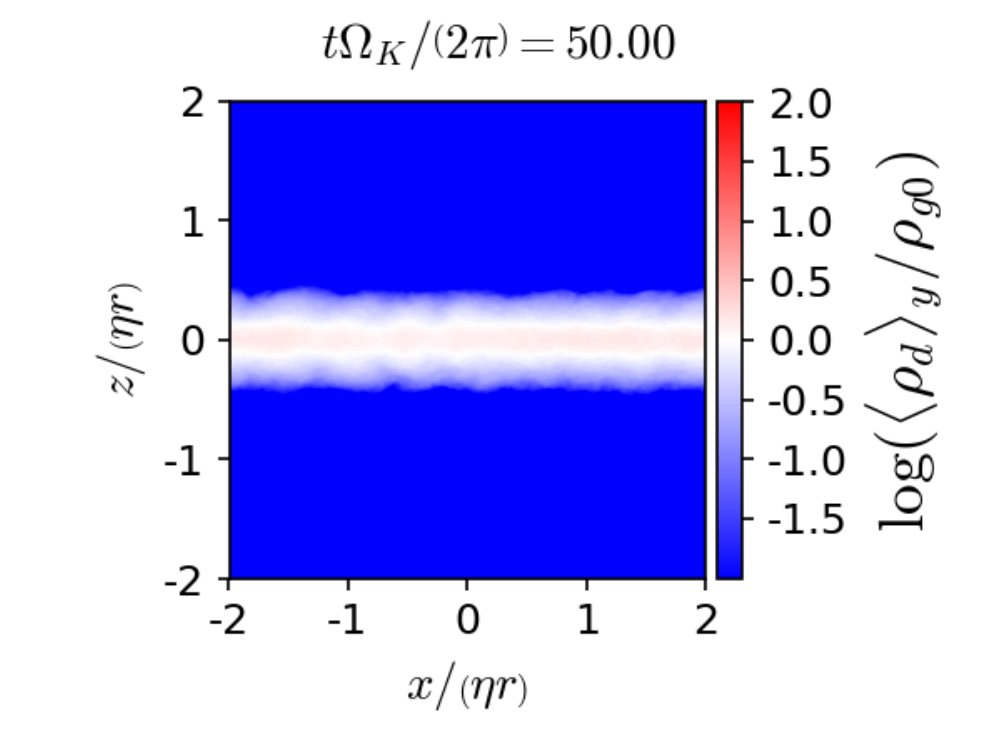}{0.45\textwidth}{(G) $\tau_s=0.01$, $\sigma_d=0.5$, $\Pi=0.025$, $Z=0.005$}
            \fig{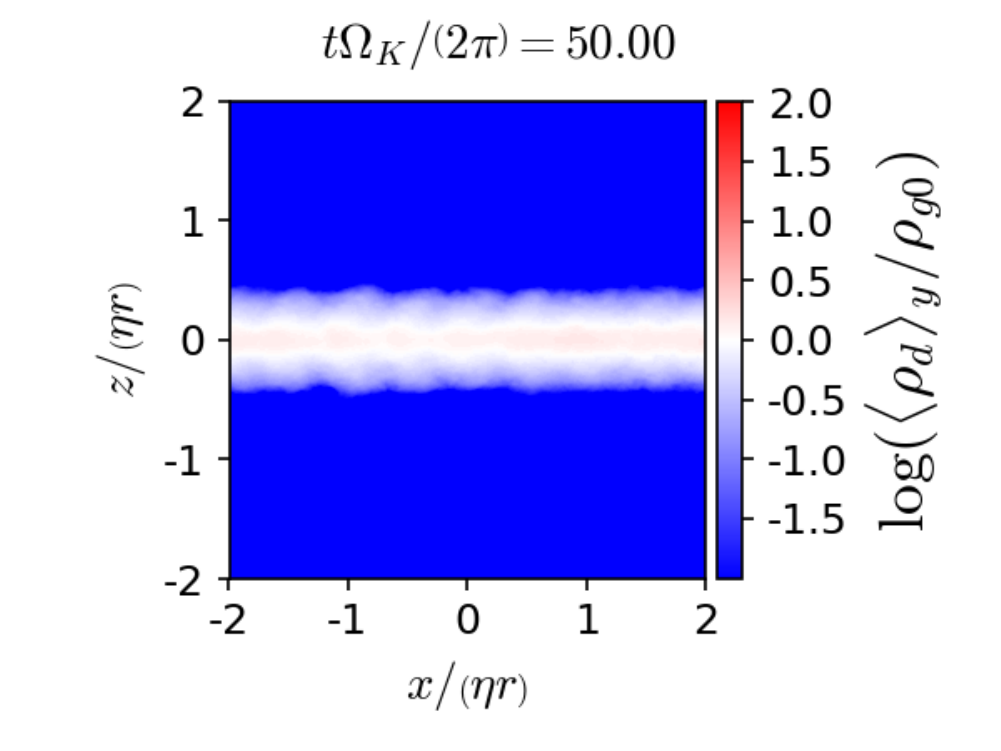}{0.45\textwidth}{(H) $\tau_s=0.01$, $\sigma_d=0.5$, $\Pi=0.05$, $Z=0.01$}
            }
\gridline{\fig{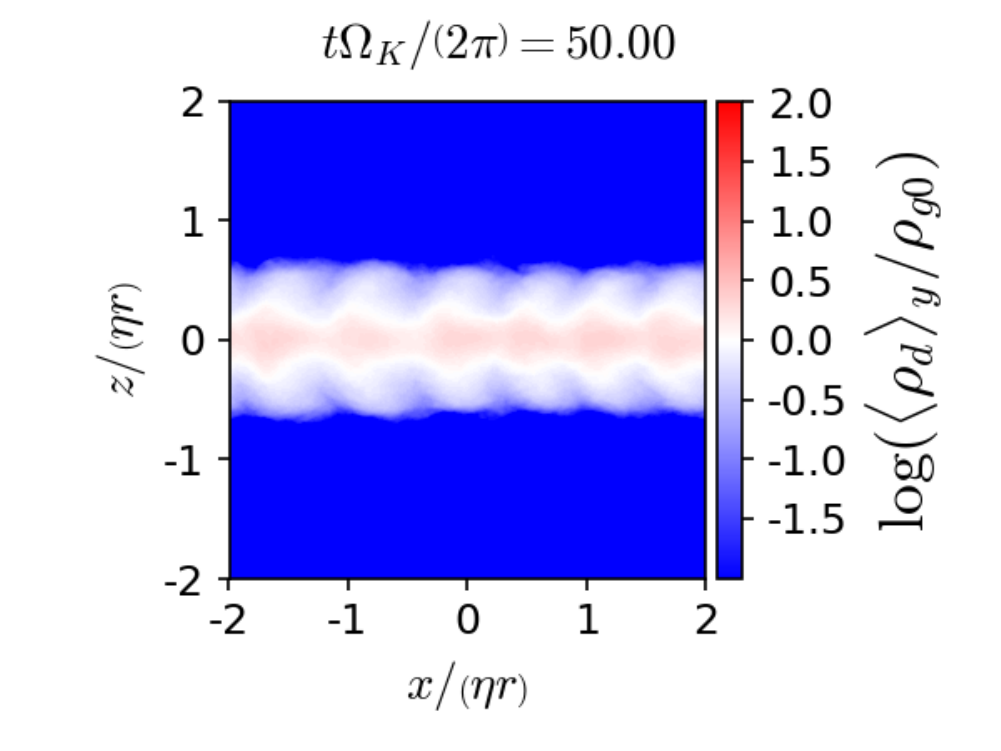}{0.45\textwidth}{(I) $\tau_s=0.01$, $\sigma_d=1.0$, $\Pi=0.025$, $Z=0.01$}
            \fig{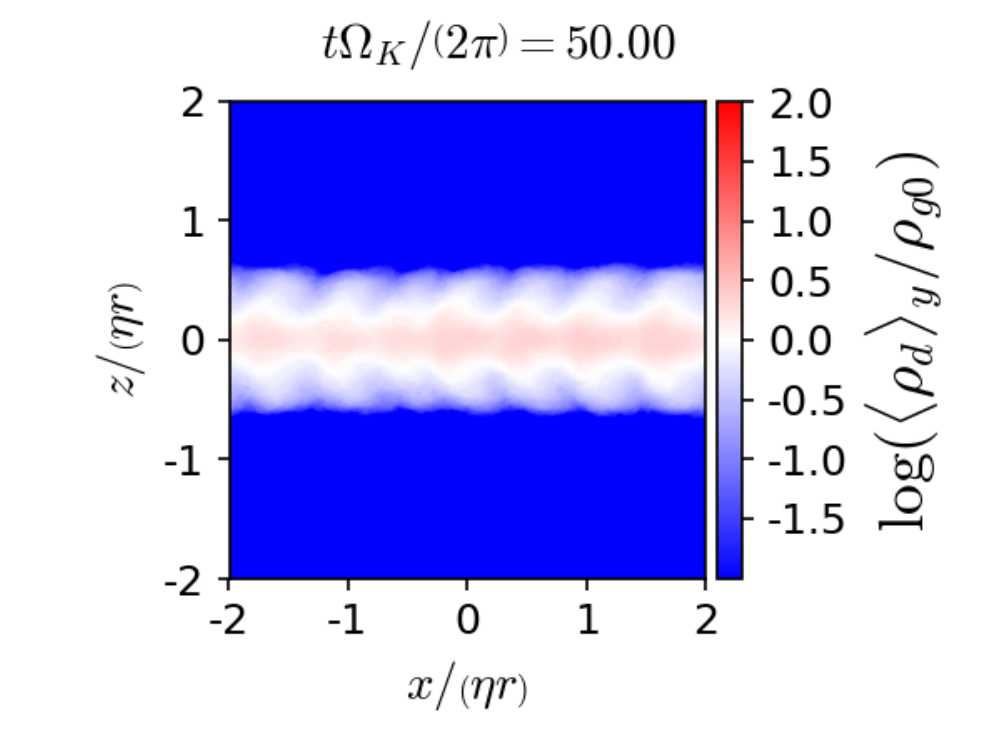}{0.45\textwidth}{(J) $\tau_s=0.01$, $\sigma_d=1.0$, $\Pi=0.05$, $Z=0.02$}
            }
\gridline{\fig{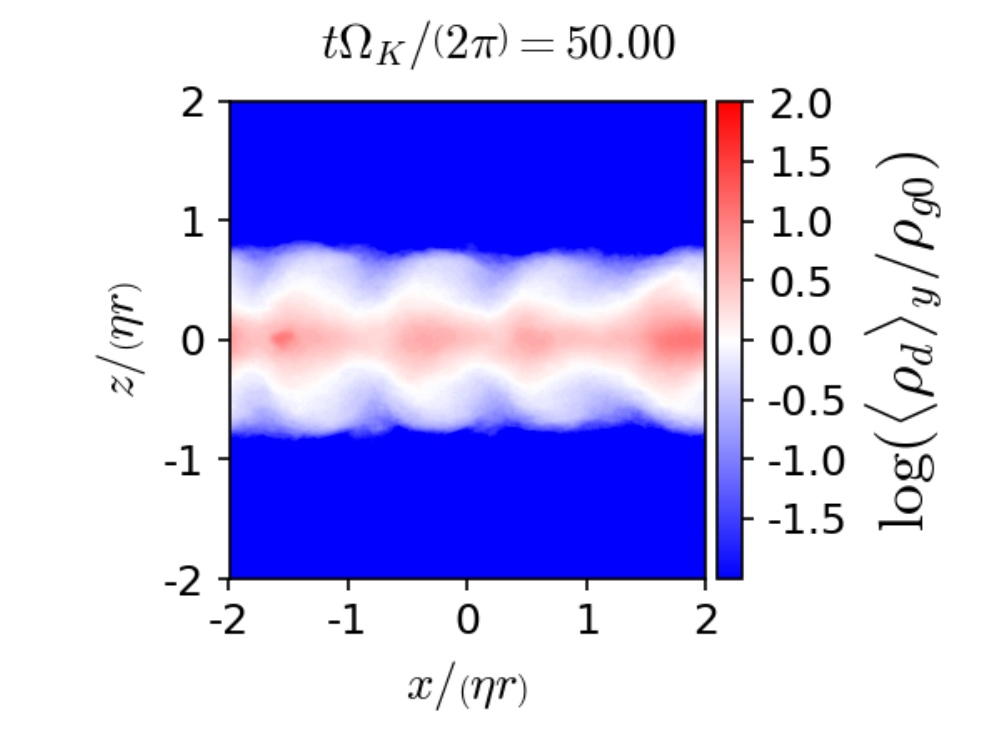}{0.45\textwidth}{(K) $\tau_s=0.01$, $\sigma_d=2.0$, $\Pi=0.025$, $Z=0.02$}
            \fig{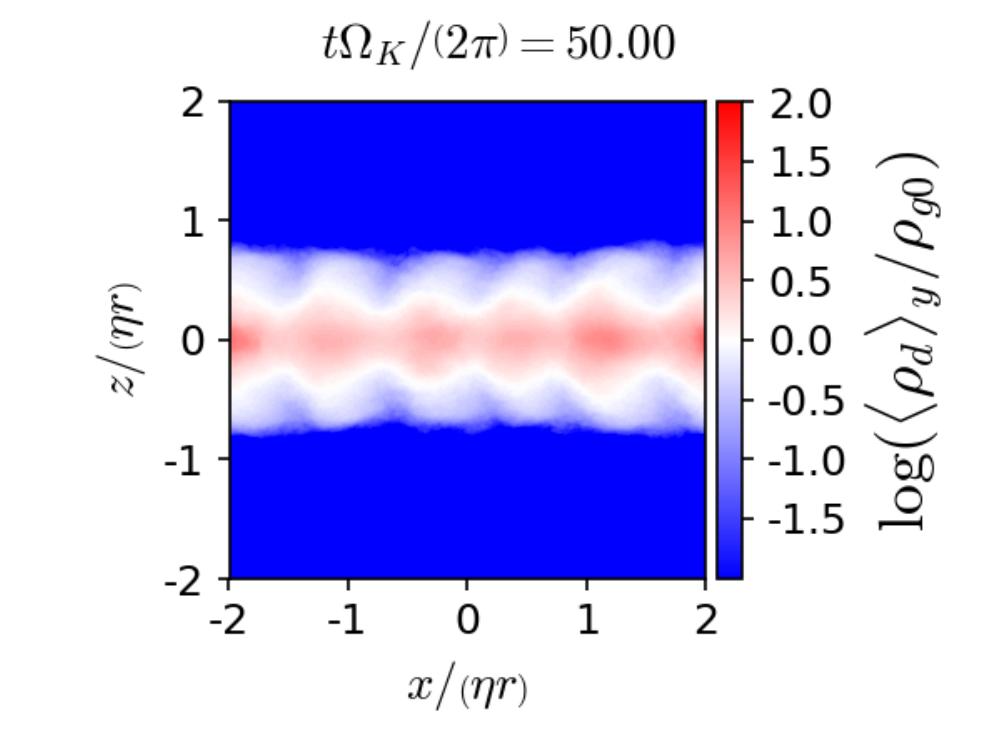}{0.45\textwidth}{(L) $\tau_s=0.01$, $\sigma_d=2.0$, $\Pi=0.05$, $Z=0.04$}
            }
\caption{
$y$-averaged dust density as a function of $(x, z)$ at $t \Omega_K / \left( 2 \pi \right) = 50$ for models (G) to (L). 
}
\end{figure*}

\begin{figure*}
\gridline{\fig{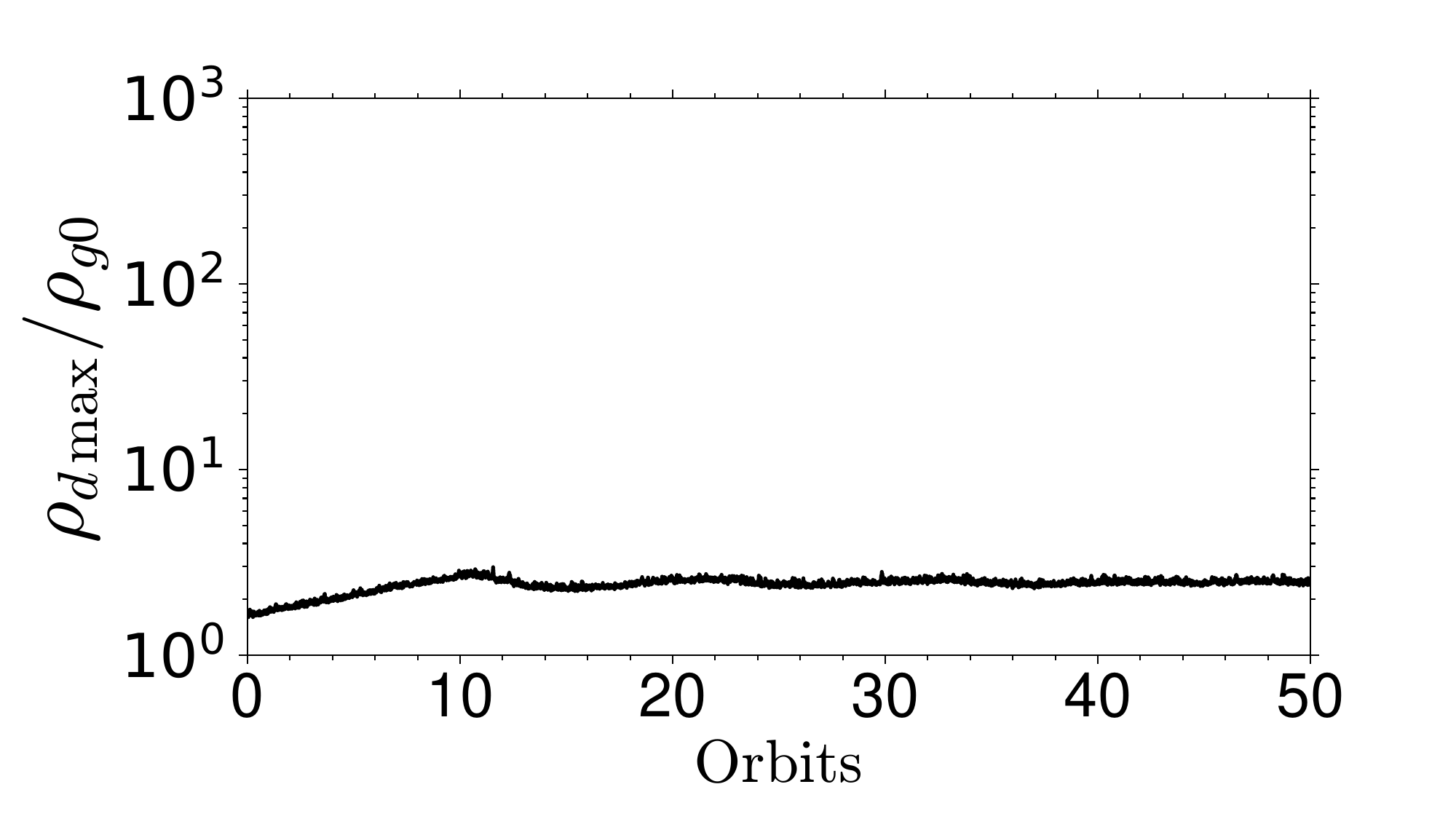}{0.45\textwidth}{(G) $\tau_s=0.01$, $\sigma_d=0.5$, $\Pi=0.025$, $Z=0.005$}
            \fig{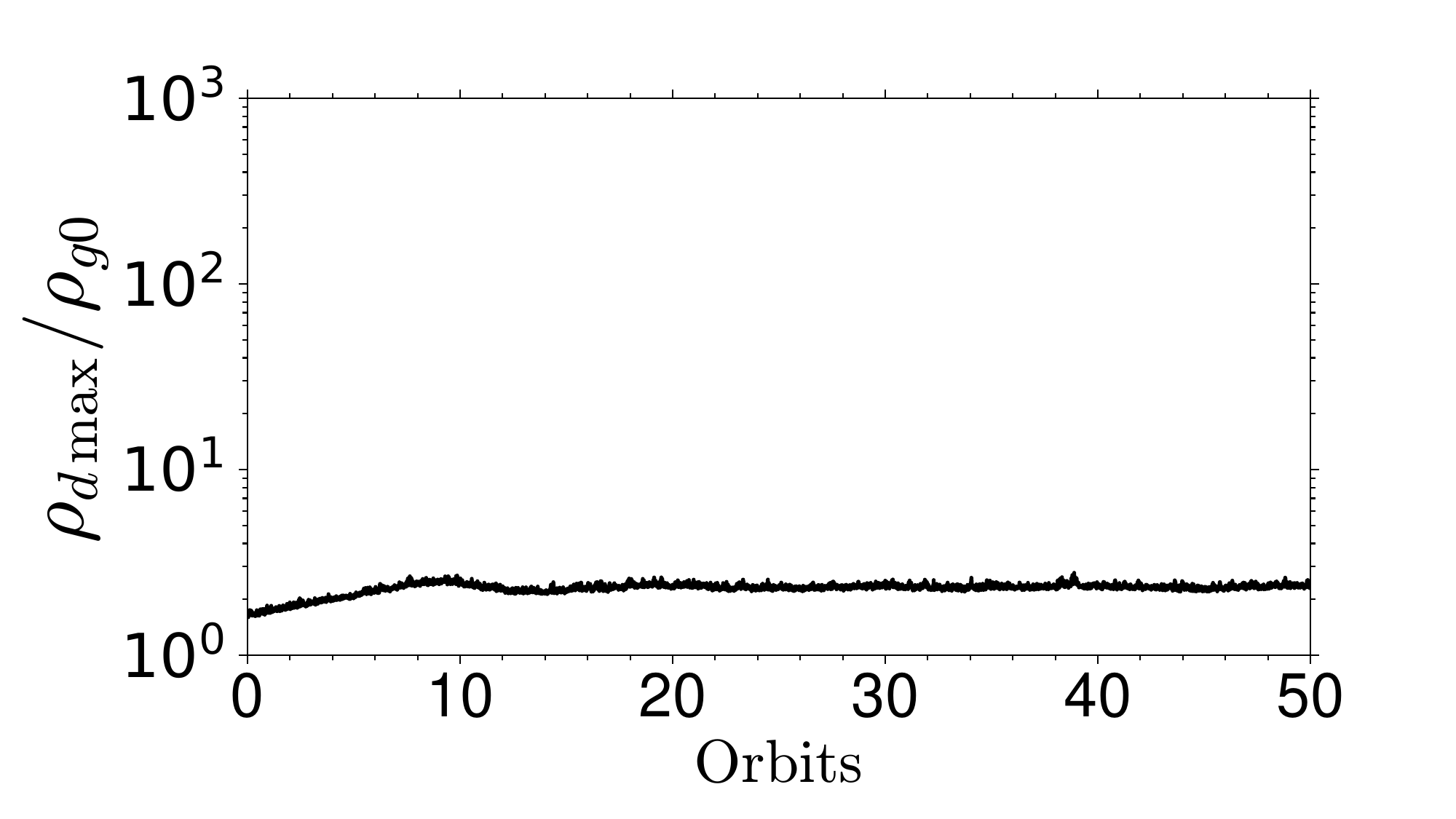}{0.45\textwidth}{(H) $\tau_s=0.01$, $\sigma_d=0.5$, $\Pi=0.05$, $Z=0.01$}
            }
\gridline{\fig{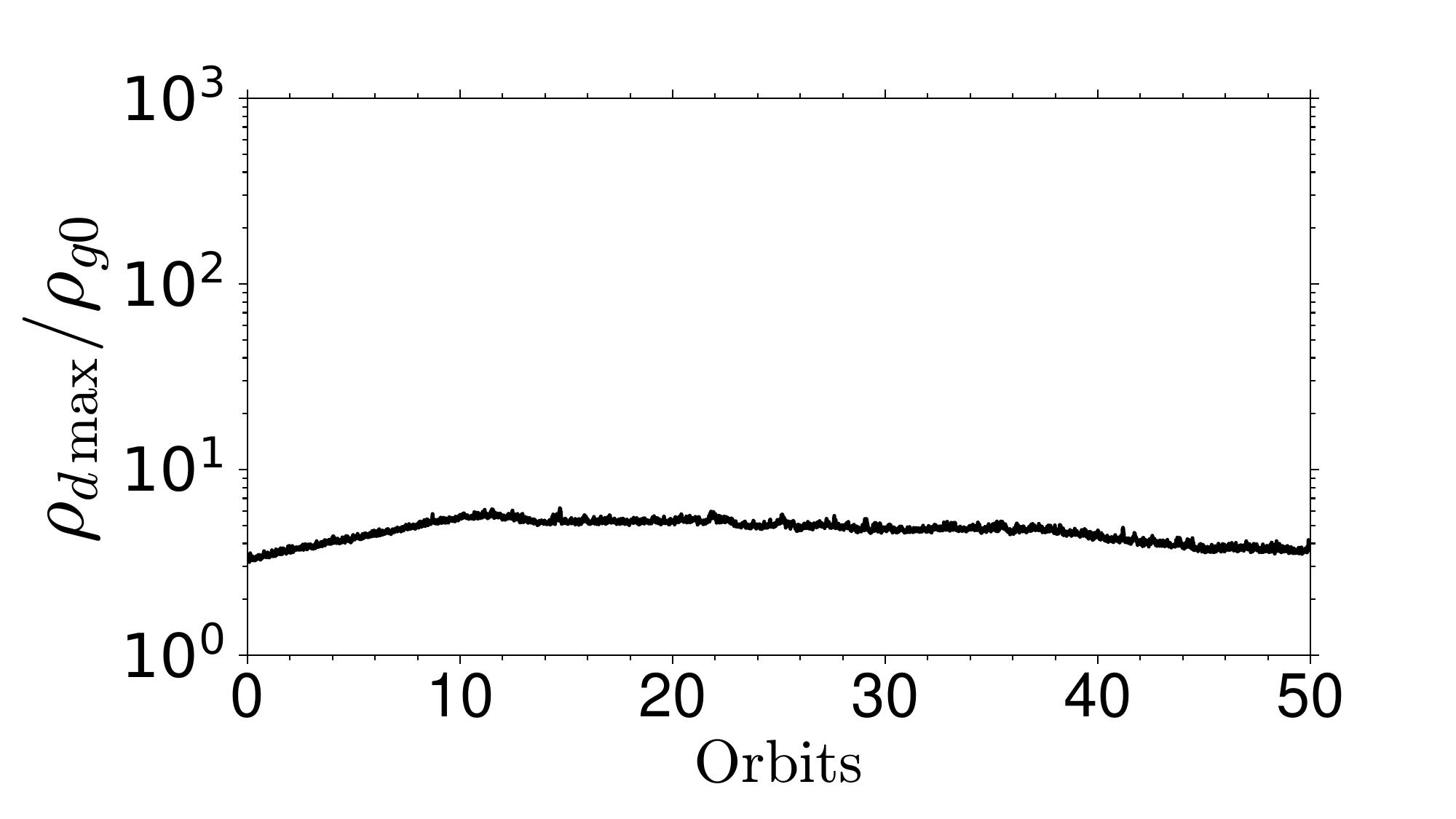}{0.45\textwidth}{(I) $\tau_s=0.01$, $\sigma_d=1.0$, $\Pi=0.025$, $Z=0.01$}
            \fig{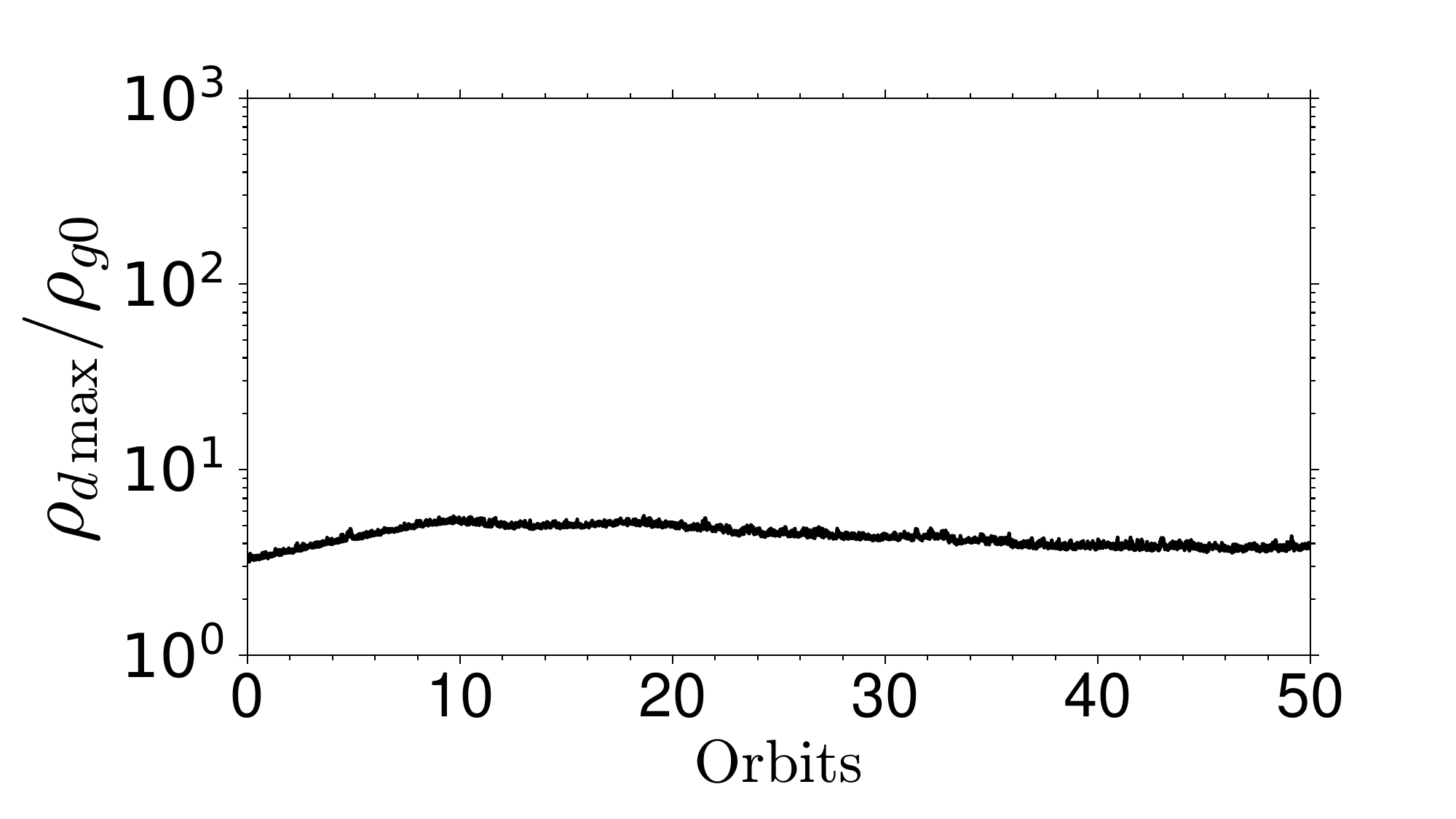}{0.45\textwidth}{(J) $\tau_s=0.01$, $\sigma_d=1.0$, $\Pi=0.05$, $Z=0.02$}
            }
\gridline{\fig{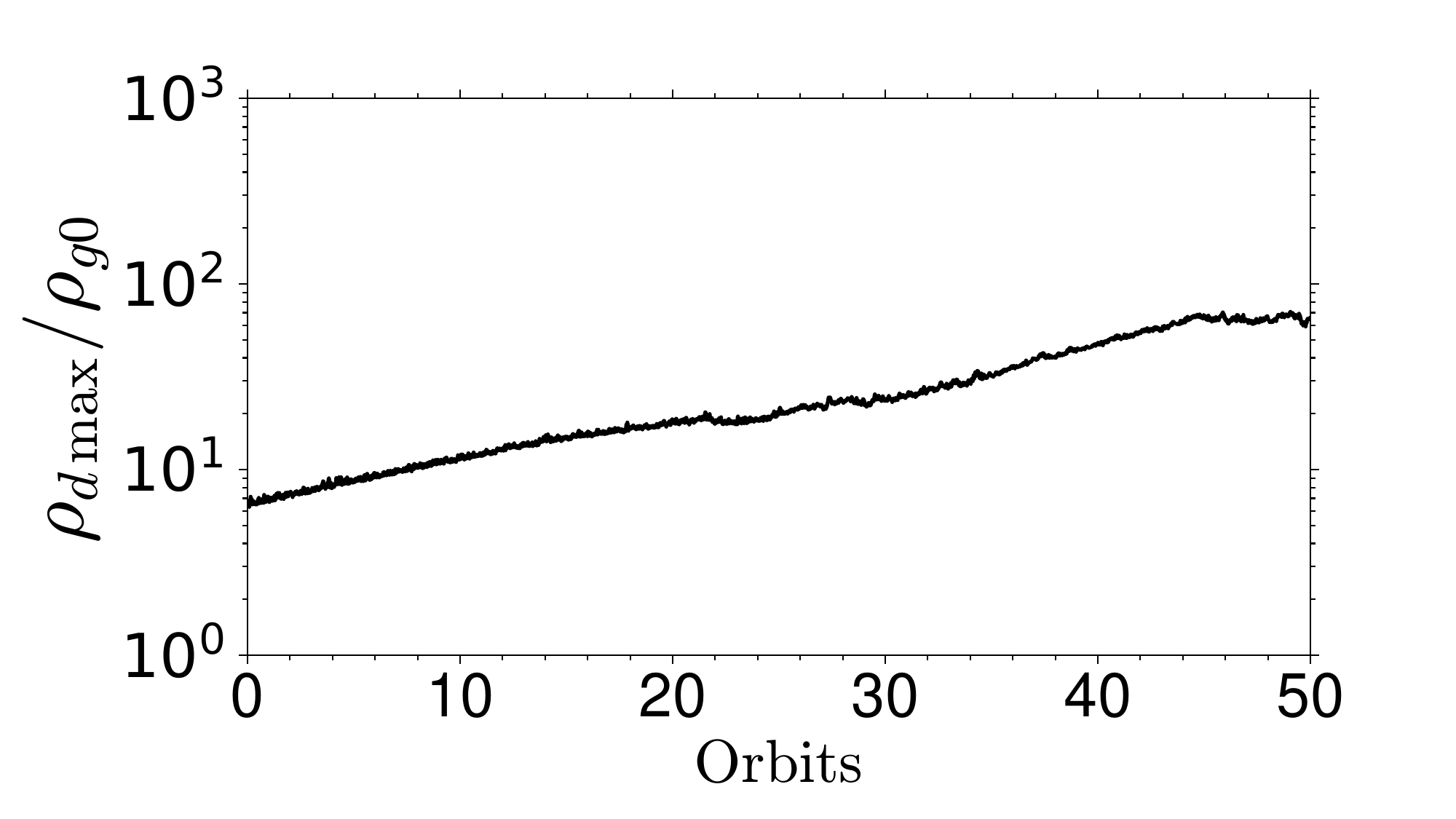}{0.45\textwidth}{(K) $\tau_s=0.01$, $\sigma_d=2.0$, $\Pi=0.025$, $Z=0.02$}
            \fig{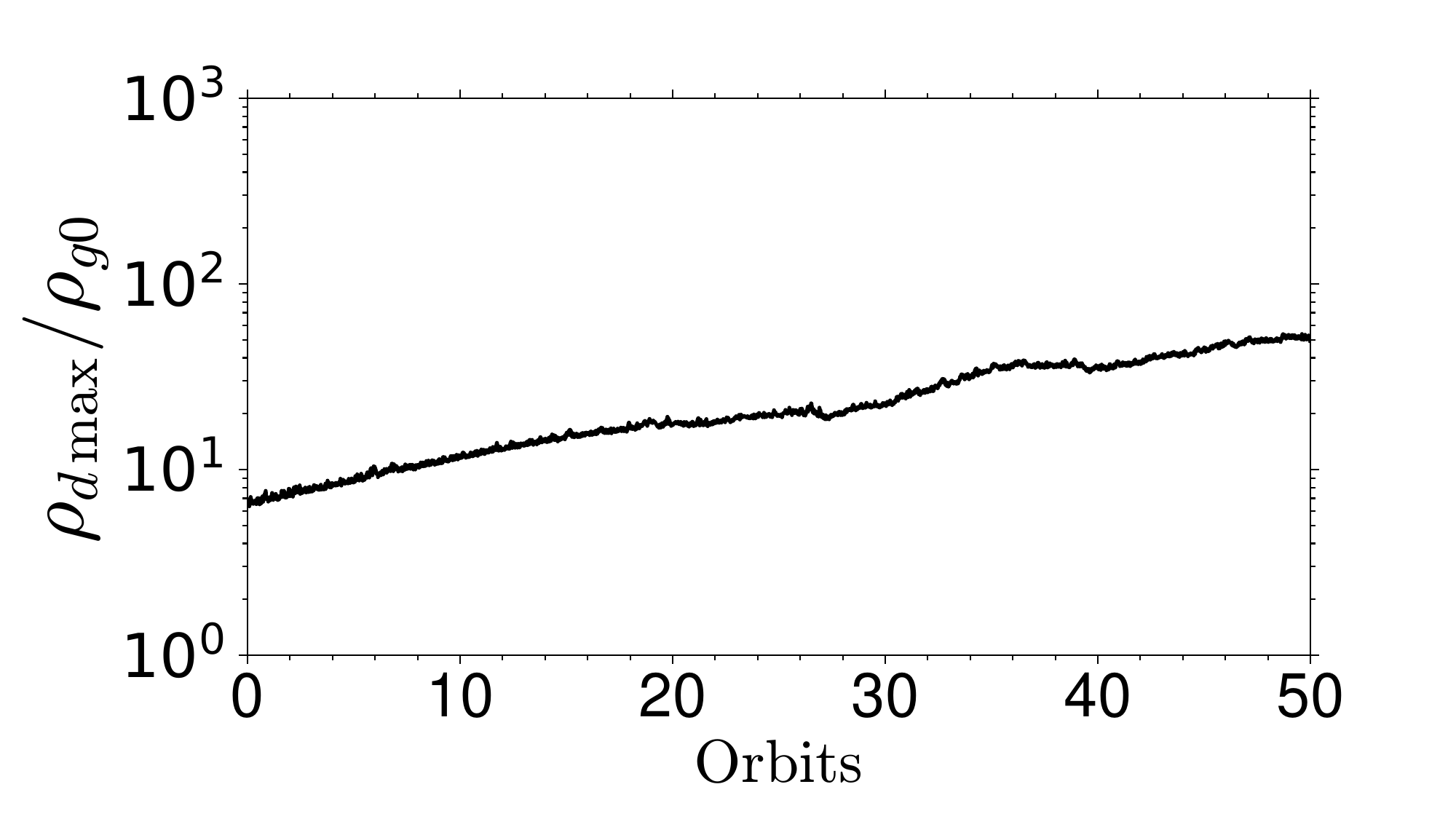}{0.45\textwidth}{(L) $\tau_s=0.01$, $\sigma_d=2.0$, $\Pi=0.05$, $Z=0.04$}
            }
\caption{Maximum dust density as a function of time (orbits) for models (G) to (L).}
\end{figure*}

\begin{figure}[ht!]
\gridline{\fig{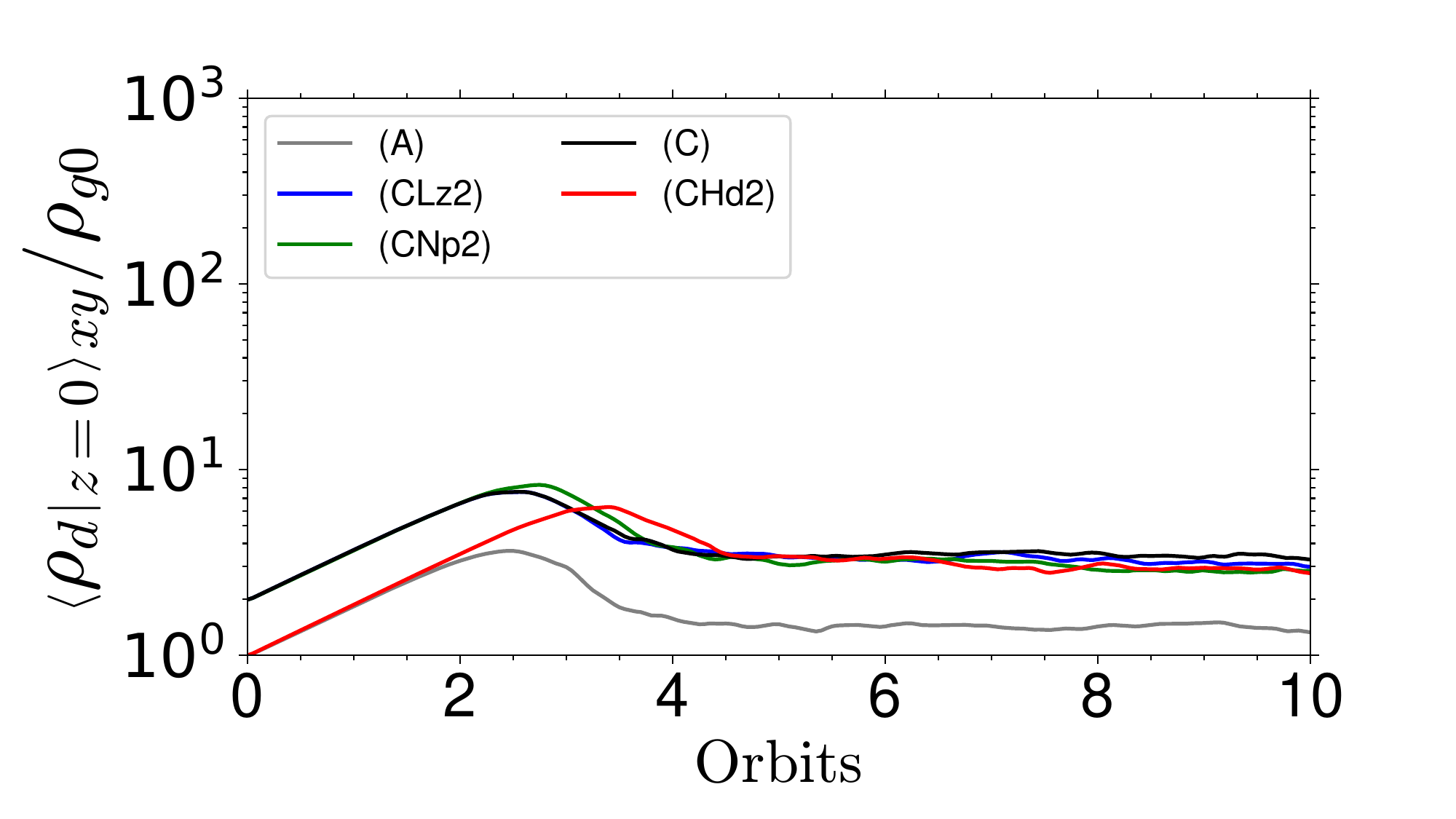}{0.45\textwidth}{(a)}
            \fig{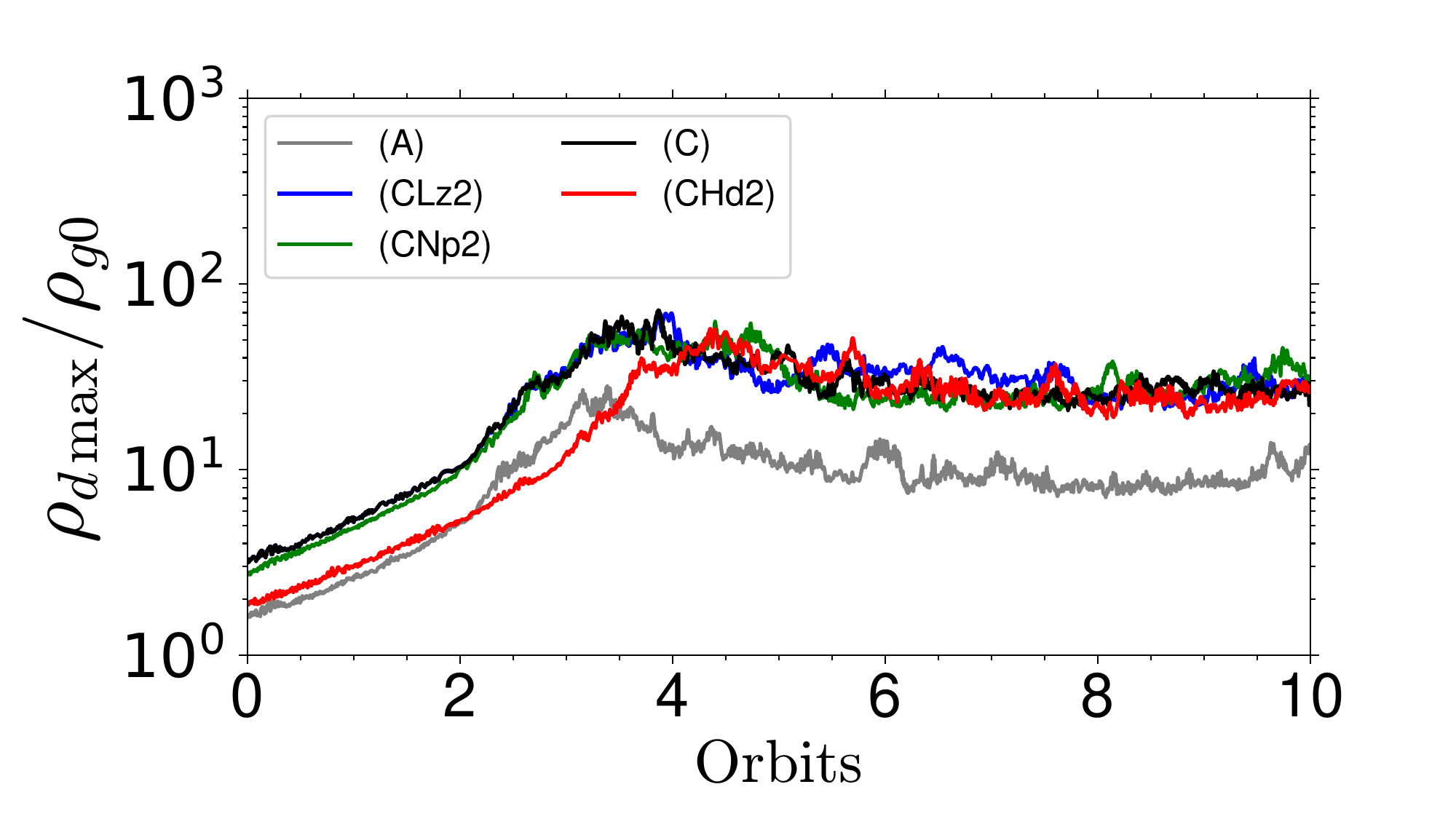}{0.45\textwidth}{(b)}}
\caption{The comparison of the time evolution of (a) the dust volume density averaged over $(x,y)$ at the midplane 
$\left< \left. \rho_d \right|_{z=0} \right>_{xy}$, 
and 
(b) the maximum dust volume density, 
$\rho_{d\,\mbox{max}}$.
Lines for models (A), (C), (CLz2), (CHd2), and (CNp2) are drawn in gray, black, blue, red, and green, respectively.} 
\end{figure}

\clearpage

\end{document}